\documentclass[useAMS,usenatbib]{mn2e}
\usepackage{times,graphicx,amsmath,amsfonts,amssymb,aas_macros,epstopdf,ulem}
\usepackage{epsfig}
\usepackage[usenames,dvipsnames]{xcolor}
\usepackage{url}
\usepackage{subfigure}

\usepackage[pdfstartview=FitV, pdftitle={The Copernicus Complexio: a high-resolution view of the small-scale dark Universe},pdfauthor={Wojciech A. Hellwing}]{hyperref}

\def\etal{{\frenchspacing\it et al.}}
\def\ie{{\frenchspacing\it i.e.}}
\def\eg{{\frenchspacing\it e.g.}}

\def\be{\begin{equation}}
\def\ee{\end{equation}}
\def\ba{\begin{eqnarray}}
\def\ea{\end{eqnarray}}
\def\Msun{h^{-1}{\rm M}_{\odot}}
\def\hmpc{h^{-1}\,{\rm Mpc}}
\def\hmpcc{h^{-3}\,{\rm Mpc}^3}
\def\hkpc{h^{-1}\,{\rm kpc}}
\def\dd{\textrm{d}}
\def\lcdm{\Lambda{\rm CDM}}

\def\ln{{\rm ln}\,}

\def\frac#1#2{{\textstyle{#1\over #2}}}

\def\simlt{\stackrel{<}{{}_\sim}}
\def\simgt{\stackrel{>}{{}_\sim}}

\newcommand{\MI}{\textsc{ms}}
\newcommand{\MII}{\textsc{ms-II}}
\newcommand{\scolor}{\textsc{color}}
\newcommand{\coco}{\textsc{coco}}
\newcommand{\aquarius}{\textsc{aquarius}}
\newcommand{\gadgetII}{\textsc{gadget2}}
\newcommand{\gadgetIII}{\textsc{gadget3}}
\newcommand{\subfind}{\textsc{subfind}}

\title[Copernicus Complexio]{The Copernicus Complexio: a high-resolution view of the small-scale Universe}
\author[Wojciech A.~Hellwing \etal]{Wojciech A. Hellwing$^{1,2}$\thanks{E-mail: pchela@icm.edu.pl}, Carlos S. Frenk$^{1}$,
Marius Cautun$^{1}$, Sownak Bose$^{1}$, 
\newauthor
John Helly$^{1}$, Adrian Jenkins$^{1}$, Till Sawala$^{1}$ and Maciej Cytowski$^{2}$\\ 
$^{1}$Institute for Computational Cosmology, Department of Physics, Durham University, South Road, Durham DH1 3LE, UK\\
$^{2}$Interdisciplinary Centre for Mathematical and Computational Modelling (ICM), University of Warsaw, ul. Pawi\'nskiego 5a, 02-106 Warsaw, Poland\\
}
\begin{document}

\date{Accepted 2016 January 23. Received 2016 January 11; in original form 2015 May 22}

\pagerange{\pageref{firstpage}--\pageref{lastpage}} \pubyear{2015}

\maketitle

\label{firstpage}

\begin{abstract}
We introduce {\it Copernicus Complexio} (\coco{}), a high-resolution
cosmological N-body simulation of structure formation in the $\lcdm{}$
model. \coco{} follows an approximately spherical region of radius $\sim
17.4\hmpc$ embedded in a much larger periodic cube that is followed at
lower resolution.  The high resolution volume has a particle mass of
$1.135\times10^5\Msun$ (60 times higher than the Millennium-II
simulation).  \coco{} gives the dark matter halo mass function over eight
orders of magnitude in halo mass; it forms $\sim 60$ haloes of galactic
size, each resolved with    about 10 million particles. We confirm the
power-law character of the subhalo mass function,
$\overline{N}(>\mu)\propto\mu^{-s}$, down to a reduced subhalo mass
$M_{sub}/M_{200}\equiv\mu=10^{-6}$, with a best-fit power-law index,
$s=0.94$, for hosts of mass $\langle M_{200}\rangle=10^{12}\Msun$.
The concentration-mass relation of
\coco{} haloes deviates from a single power law for masses
$M_{200}<\textrm{a few}\times 10^{8}\Msun$, where it flattens, in
agreement with results by Sanchez-Conde et al. The host mass
invariance of the reduced maximum circular velocity function of
subhaloes, $\nu\equiv V_{max}/V_{200}$, hinted at in previous
simulations, is clearly demonstrated over five orders of magnitude
in host mass. Similarly, we find that the average, normalised radial
distribution of subhaloes is approximately universal
(i.e. independent of subhalo mass), as previously suggested by the
Aquarius simulations of individual haloes.  Finally, we find that
at fixed physical subhalo size, subhaloes in lower mass hosts
typically have lower central densities than those in higher mass
hosts.
\end{abstract}

\begin{keywords}
cosmology: theory, dark matter - methods: numerical
\end{keywords}

\section{Introduction}
\label{sec:intro}

Since its introduction over thirty years ago, the cold dark matter
(CDM) model of structure formation
\citep{Peebles1982,Davis1985,Bardeen1986} has been extensively
investigated theoretically and tested with an impressive array of
observational data.  According to this, the now standard, model of
cosmogony, galaxy formation is driven by the evolution of the dark
matter (DM) haloes in which the galaxies reside. It is into these
haloes that gas cools and condenses, becomes unstable and fragments
into stars, leading to the formation of galaxies
\citep{WhiteRees1978,WhiteFrenk1991}.  This basic picture has been
elaborated in detail using simulations and semi-analytic models and it
has largely been confirmed by countless observations \citep[see
\eg][for a recent review]{FrenkWhite2012}. Thus, DM haloes are the
fundamental non-linear building blocks of cosmic structure
\citep{Merging_halgal_Kauffmann1993,ColeLacey1996} and understanding
their properties, abundance and spatial distribution has been
a subject of extensive theoretical study over the past thirty years.

The formation and evolution of DM haloes and of the galaxies residing
within them is modelled in the context of the background cosmological
model that describes the expansion and growth history of the
Universe. The last 20 years have seen the emergence of the ''Lambda
Cold Dark Matter'' ($\lcdm$) model, which combines a flat
Friedmann-Lema\^\i tre model with a cosmological constant - $\Lambda$
-, responsible for the late-time accelerated expansion of the
Universe, with the CDM paradigm, in which thermally cold relic
elementary particles constitute the majority of non-relativistic
matter and govern the growth of cosmic structures.  The predictions of
the $\lcdm$ model have been tested to a high precision on linear
scales, from the very early universe
\citep[\eg][]{WMAP9,Planck2013,Planck2015} to large cosmic scales
\citep[\eg][]{BAO_2dF,BAO_SDSS,BAO_SDSS7,BOSS,VIPERS,GAMMA}, yielding
very good agreement with observations. To further test and constrain
the current model, one needs to study its predictions down to smaller
scales, extending significantly into the non-linear regime of
structure formation.

N-body simulations represent the most widely used and convenient
method of exploring the highly non-linear regime of cosmic structure
formation. Starting from a set of initial conditions, the numerical
simulations follow the formation and evolution of structures from an
early epoch down to present day. Motivated by the fact that DM
represents most of the matter in the Universe and because of the
relatively simple physics of collisionless DM particles, DM-only
simulations represent the most widely used category of numerical
simulations.  When designing a cosmological N-body experiment, one is
concerned by two major factors. Ideally, one would like to simulate a
region of the universe that is as large as possible to get a
representative census of the structures encompassed within it.  On the
other hand, one would also want very high mass resolution, to be able
to resolve accurately even the smallest cosmologically relevant
objects. Unfortunately, due to limited computational resources, these
two requirements are in conflict, which implies that various
compromises need to be made when designing a numerical simulation.  So
far, the biggest efforts were focused into two, somewhat complementary
approaches.  The first is represented by simulations like Millennium
(\MI{}) \citep{MS}, Millennium II (\MII{}) \citep{MS2}, Millennium XXL
(MXXL) \citep{MXXLAngulo2012}, Bolshoi \citep{Bolshoi}, MultiDark
\citep{Prada2012}, Horizon Run I-III \citep{Kim2009,Kim2011},
Horizon-$4\pi$ \citep{Prunet2008,Teyssier2009}, MareNostrum Universe 
\citep{Gottloeber2006}, Jubilee project \citep{Watson2014}, Coyote Universe
\citep{Heitmann2010}, DEUS simulation \citep{Alimi2012,Rasera2014}
or MICE suite \citep{Fosalba2015}. These follow structure formation in a large
cosmological volume at the expense of having a medium or a low mass
resolution.  Such simulations provide the formation histories for a
very large number of medium- and high-mass DM haloes, but do not
necessary resolve all the details relevant for galaxy formation.  On
the other side we have N-body simulations like the \aquarius{} project
\citep{Aquarius}, the Via Lactea \citep{ViaLactea}, the Phoenix
project \citep{Phoenix}, CLUES \citep{Gottloeber2010}
and the ELVIS suite \citep{Elvis} that are
characterised by a very high mass and force resolution but are limited
to very small cosmic volumes. These give a very detailed picture of
galaxy- and cluster-size haloes, but do so only for a very limited
number of objects, which makes their results sensitive to small number
statistics, and are unable to capture the full interconnection between
small (DM haloes) and large (the cosmic web) cosmic scales.

The recent years have seen a lot of attention focused on obtaining detailed histories for a 
large number of Galatic-size DM haloes. This is because our own Milky Way (MW)
Galaxy together with the Local Group (LG) galaxies, thanks to their direct proximity,
constitute an important test-bed for cosmic structure formation theories. Thanks to an ever growing
accuracy of astronomical observations, we are presented with a very detailed picture 
of our nearest cosmic neighbourhood. The past decade has brought an impressive 
amount of data on the MW, Andromeda and their satellites, as well as on other small
members of the LG \citep[\eg][]{Belokurov2006,Belokurov2007,Belokurov2014,Pandas,Koposov2008,McConnachie2012}.
These data have led to a number of apparent discrepancies between the predictions of numerical simulations
and observations, which are collectively known as the ``$\lcdm$ small-scale crisis''. 
The ``missing satellites problem'' \citep{Merging_halgal_Kauffmann1993,Klypin1999,Moore1999}
was among the first to be recognised. Here the tension arises due to the fact that 
dissipationless numerical simulations predict many more small DM satellites (clumps or subhaloes)
in a Galactic-size halo than the actual number of observed MW satellites. 
One of the most favoured solution to this problem predicts that below a certain mass-scale 
the majority of DM satellites have failed to host luminous galaxies. This is due to baryonic physics, 
related to hydrodynamic, energy feedback processes and the reionisation, that depletes the cold gas from small mass haloes, 
thus preventing star formation and rendering these objects dark 
\citep[for the most recent results see \eg][]{Eagle,2014Natur.509..170B,Illustris,Sawala2014b,Sawala2014c}.

Another small-scale $\lcdm$ discrepancy, emphasised in recent years by \cite{2011MNRAS.415L..40B},
is the so-called ``Too Big Too Fail'' problem. It is due to the inconsistency between the internal 
kinematics of the observed 11 classical dwarf MW satellites and the distribution of kinematic parameters
inferred for the most massive satellites of MW-size hosts in the \aquarius{} simulation suite 
\citep{2012MNRAS.422.1203B}.
Recently, a similar claim was made also for the field dwarf galaxies found in the LG \citep{Garrison-Kimmel2014}.
This discrepancy has various possible solutions, being a possible manifestation of highly non-linear
and stochastic baryonic physics in low mass haloes \citep{Sawala2014a} and the impact of stellar feedback
on DM density profiles \citep[see \eg][]{Pontzen2012,Onorobe2015,Brook2015}.
Others have also shown that the problem can be largely alleviated 
when the mass of the MW (LG) is sufficiently low \citep[\eg][]{Wang2012,Cautun2014a}.

Finally, the recent discovery that a subset of Andromeda satellites are distributed in a thin plane \citep[\eg][]{McConnachie2006,Ibata2013}
and their radial velocity components show some degree of a coherent co-rotation,
together with the previously known thin polar disk-like distribution of the MW satellites \citep[\eg][]{Metz2008,Pawlowski2013},
were postulated by some authors to also present a challenge for the $\lcdm$ paradigm.

It is important to note that many of these apparent points of tensions were derived
from comparisons with a rather small number of host haloes. In particular, obtaining sufficient resolution to study
the internal properties of subhaloes in a MW-size host necessitates the use of
ultra-high resolution simulations, which are limited to very small volumes and only a handful of central host haloes.
Given such a limited sample of host haloes, the resulting satellite populations may be prone to halo-to-halo scatter,
which is intrinsic to hierarchical models for structure formation. MW-size haloes are characterised
by variety of evolutionary histories and large-scale structure environments in which these systems evolve.
These are important factors that need to be properly evaluated and understood before 
claiming any potential discrepancies between $\lcdm$ galactic-scale predictions and the MW and LG
observations. For example, proper cosmological-volume simulations could help us determine to which
extent our own Galaxy, its DM halo and the satellite system are rare or special within the $\lcdm$ paradigm.
This concerns one of the core assumptions of modern cosmology, the {\it Copernican Principle},
according to which a MW-based observer is not privileged in the sense that we can observe a fair
sample of the Universe.

In this paper we introduce the {\it Copernicus Complexio} (the Copernicus Conundrum; hereafter \coco{}), which
is a DM-only simulation tailored for the study of a statistically significant sample of well resolved 
MW-size haloes and their satellites. The simulation follows a hybrid 'zoom-in' approach, similar to the one 
adopted in the GIMIC simulation suite\citep{GIMIC} (\textit{Galaxies-intergalactic medium interaction calculation}),
with a high-resolution
region of radius ${\sim}17\hmpc$ embedded within a much larger box resolved at low-resolution. The large
volume of the high-resolution region contains around 60 MW-size haloes and their satellite populations, 
resolved at a resolution close to that of the \aquarius{} level 3 simulations. This is more than sufficient to properly 
capture the internal structure and properties of subhaloes hosting faint MW satellites, attaining
at the same time a good statistical sample of DM hosts of various masses located in diverse environments.
In addition, the simulation contains a very large number of well-resolved lower mass haloes, whose properties 
are studied here for the first time with such good statistics. 

In this paper we introduce the new \coco{} simulation and present the first-stage analysis of its results. Section 2
presents our selection of the high-resolution region and 
gives details on the numerical and cosmological set-up. The results on DM halo abundances, formation times 
and internal density profiles are presented in Section 3. In Section 4 we study in detail the populations
of satellite subhaloes, including their mass and velocity functions, radial distributions, internal
kinematics and effects of host-induced tidal stripping. We give our concluding remarks in the Section 5.

\section{The Copernicus Complexio cosmological simulation}
\label{sec:simulation}

\begin{figure}
  \includegraphics[width=85mm]{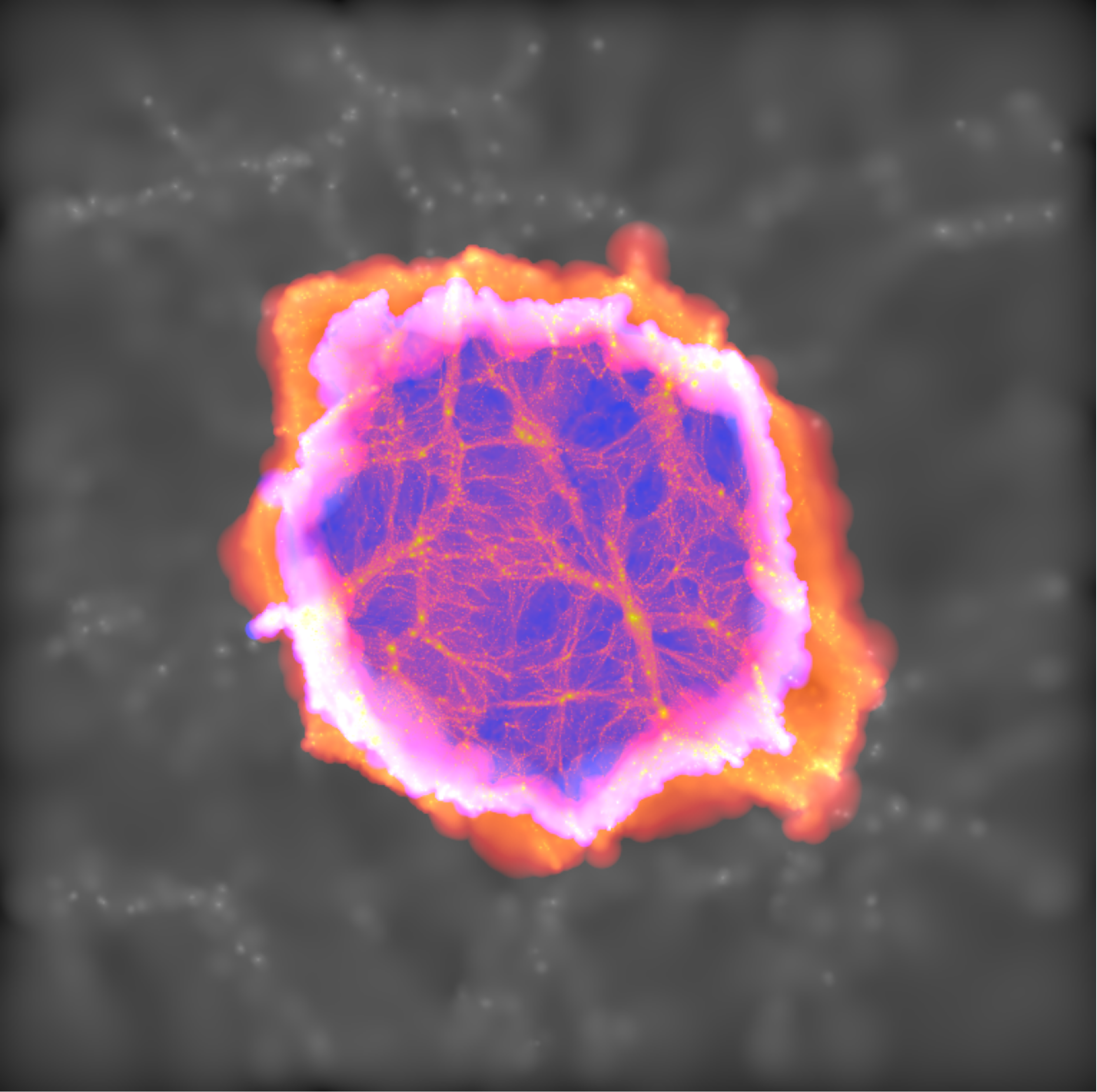}
  \caption{Projected density along the z-axis of a
    $70.4\times70.4\times1.5\hmpc$ slice centred on the middle of the
    \coco{} simulation at redshift $z=0$.  The various colours show
    the density at different resolution levels: lowest resolution
    (grey), medium resolution (orange and purple) and high-resolution
    (blue to yellow).  Note the amazing level of the cosmic web
    details seen inside the high-resolution region.}
\label{fig:coco_dens_1}
\end{figure}
The \coco{} simulation was designed with the goal of resolving the formation and evolution of MW-size haloes and 
their subhaloes in a representative cosmological volume. This prompted the use of a zoom-in simulation \citep{zoom1,Frenk1996,GIMIC,zoom2} that 
captures in very great detail the evolution of a selected region, which in turn is embedded within a larger 
cosmological volume that is simulated at low-resolution. The role of the latter is to produce the correct large scale modes 
and tidal fields inside the high-resolution region. Starting from a low-resolution simulation, the high-resolution 
volume was selected by optimizing the number of Galactic-mass haloes that could be resolved given the available 
computational resources. 

Randomly selecting the high-resolution region can result in it containing one or more rich clusters. Such 
massive objects would dominate the computational time required for the whole \coco{} simulation, leading to a 
wastage of resources, since we are primarily interested in MW and lower mass objects. 
To avoid unnecessary computations, but keeping in mind that we want to simulate a fair-sample of the Universe, 
possibly close to the observed Local Volume, we have selected a region that
satisfies the following criteria:
\begin{enumerate}
 \item there are no cluster-mass haloes ($M\simgt5\times 10^{13}\Msun$) inside the zoom-in region,
 \item there are no massive cluster haloes ($M\simgt5\times 10^{14}\Msun$) within $5\hmpc$ of 
the zoom-in boundary,
 \item the mass function of MW-mass haloes ($M\sim10^{12}\Msun$) is as close as possible 
to the universal mass function.
\end{enumerate}

\subsection{Cosmological and numerical parameters}
The \coco{} simulation follows structure formation in a high-resolution region that is approximately 
a sphere of radius $\sim 17.4\hmpc$ ($2.2\times10^4\hmpcc$ in volume) embedded within 
a $70.4\hmpc$ low-resolution periodic box, as illustrated in Fig.~\ref{fig:coco_dens_1}. 
It uses a \textit{Wilkinson Microwave Anisotropy Probe} 
(WMAP) - seventh year result  cosmogony \citep{WMAP7} with the following cosmological parameters:
\ba
\Omega_{m0}=0.272\,,\,\,\Omega_{\Lambda0}=0.728\,,\,\,\Omega_b=0.04455\,,\nonumber\\
\Omega_k=0\,,\,\,h=0.704\,,\,\,\sigma_8=0.81\,,\,\,n_s=0.967\,.
\label{eqn:wmap7-parameters}
\ea
The high resolution region consists of ${\sim}12.9$~billion particles of 
mass $1.135\times10^5\Msun$ and of ${\sim}510$~million medium and low resolution
particles that have progressively larger masses. The Plummer equivalent force 
softening was chosen to increase from a value of $0.23\hkpc$ for the high resolution
particles to a value of $23\hkpc$ for the lowest resolution level.

The high-resolution region was selected from a lower resolution
version of the \coco{} simulations that we refer to as {\it COpernicus 
complexio LOw Resolution} (\scolor{}).  The \scolor{} simulation
has the same corresponding initial phases as \coco{} but is set-up
with  $1620^3$ DM particles uniformly distributed throughout the whole
$70.4\hmpc$ periodic box. It has a mass and force resolution of $\sim
6.2\times 10^6 \Msun$ and $1\hkpc$ respectively, which is exactly the
same as in the \MII{}. While the \scolor{} box size
is relatively small, the lack of very large-scale modes has little impact
on the internal properties of galactic and smaller mass haloes
\citep[for more details see][]{Power2006}.

The selection of the \coco{} zoom-in region was performed by generating
a large number of randomly placed spheres of radius $17\hmpc$ in the \scolor{} volume at $z=0$.
We discarded any volume not fulfilling criteria \textit{(i)} and \textit{(ii)} given 
in Section \ref{sec:simulation}. From the remaining volumes, we selected the 
one whose halo mass function in the range, $M\lesssim10^{12}\Msun$, showed the closest match to the 
universal mass function. Following this, the initial conditions
of the zoom-in region were generated using the same method as in the \aquarius{} and the
\textsc{gimic} \citep{GIMIC} projects. The particles from the selected $z=0$ volume
were traced back to their Lagrangian positions. The Lagrangian volume was divided 
in $256^3$ regular cells and each cell occupied by one or more of these particles was
classified as high resolution. The remaining cells were classified as medium or low resolution 
cells depending on the distance to the nearest high resolution cell. Each high
resolution cell is filled with a periodic glass distribution of $24^3$ particles,
while the medium to low resolution cells were sampled with progressively fewer particles.
Higher frequency power was added to the resulting particles down to the Nyquist frequency
while making sure that the lower frequency modes were the same as in the \scolor{} simulation. 

The initial conditions (initial positions and velocities of all
particles) for the \coco{} simulation were set at $z=127$ using second
order Lagrangian perturbation theory using the method of
\cite{Jenkins2lpt}. The initial phases for both the \coco{} and
\scolor{} are taken from the public multi-scale Gaussian white noise
field called Panphasia, and are published in Table~6 of
\cite{Panphasia} under the alternative name of the `DOVE'
simulation. The \coco{} initial conditions differ from \scolor{} by a
uniform spatial translation so that the coordinate origin in \coco{}
is located at the coordinates $(7,16,44)\hmpc$ within the \scolor{}
simulation.  This translation places the high resolution region of \coco{}
at the centre of the simulation volume.

\begin{table*}
\begin{minipage}{\textwidth} 
\begin{center}
\caption{Details of the simulations used and described in this work along with a list of some other large simulations 
frequently used in the literature.
$V_{\rm box}$ gives the total comoving volume of the simulation, $N_p$ is the number of N-body particles, 
$\varepsilon$ denotes the Plummer-equivalent force softening, expressed
in comoving units, $m_p$ is the mass resolution.
% and $f_{halo}$ marks the fraction of mass in FOF haloes resolved with 20 or more particles (at $z=0$).  
}
\label{tab:sims_details}
\begin{tabular}{@{}lcccccc}
\hline\hline
Name & $V_{\rm box}$ [$\hmpcc$] & $N_p$ & $\varepsilon$ [$\hkpc$] & $m_p$ [$\Msun$] & cosmology & reference\\
\hline
\coco{}\footnote{Here we only consider the high-resolution region.\label{fnote}} & ${\sim}2.2\times 10^4$ & ${\sim}2344^3$\footnote{Actual particle number, $N_p=12,876,807,168$.} & 0.23 & $1.135\times10^5$ & WMAP7 & this work\\
\scolor{} & $3.5\times 10^5$ & $1620^3$ & 1.0 & $6.19\times10^6$ & WMAP7 & this work\\
Millennium-II & $1.0\times 10^6$ & $2160^3$ & 1.0 & $6.89\times10^6$ & WMAP1 & \cite{MS2}\\
Millennium  & $1.3\times10^8$ & $2160^3$ & 5.0 & $8.61\times10^8$ & WMAP1 & \cite{MS}\\
\aquarius{} lvl. 3\textsuperscript{\ref{fnote}} & ${\sim}1.1\times 10^2$ &  ${\sim}530^3$\footnote{Actual particle number, $N_p=148,285,000$.} & $0.12$ & ${\sim}5\times10^4$ & WMAP1 & \cite{Aquarius}\\
Via Lactea (LR)\textsuperscript{\ref{fnote}}& ${\sim} 5.5\times 10^2$ & ${\sim} 402^3$ & 0.378 & $4.11\times 10^{5}$ & WMAP3 & \cite{ViaLactea}\\
Horizon-$4\pi$ & $8.0\times10^9$ & $4096^3$ & 7.6 & $7.7\times 10^{9}$ & WMAP3 & \cite{Teyssier2009}\\
Bolshoi         & $1.6\times10^7$& $2048^3$ & 1.0 & $1.35\times10^8$&combination\footnote{X-ray clusters+WMAP5 +SN+BAO.}& \cite{Bolshoi} \\
Horizon Run 3 & $1.3\times10^{12}$ & $7210^3$ & 150 & $2.44\times 10^{11}$ & WMAP5 & \cite{Kim2011}\\
Jubilee  & $2.2\times10^{11}$ & $6000^3$ & no data & $7.49\times 10^{10}$ & WMAP5 & \cite{Watson2014}\\
MICE & $2.9\times10^{10}$ & $4096^3$ & 50 & $2.93\times 10^{10}$ & WMAP5 & \cite{Fosalba2015}\\
\hline
\end{tabular}
\end{center}
\end{minipage}
\end{table*}

\begin{figure*}
  \includegraphics[width=175mm]{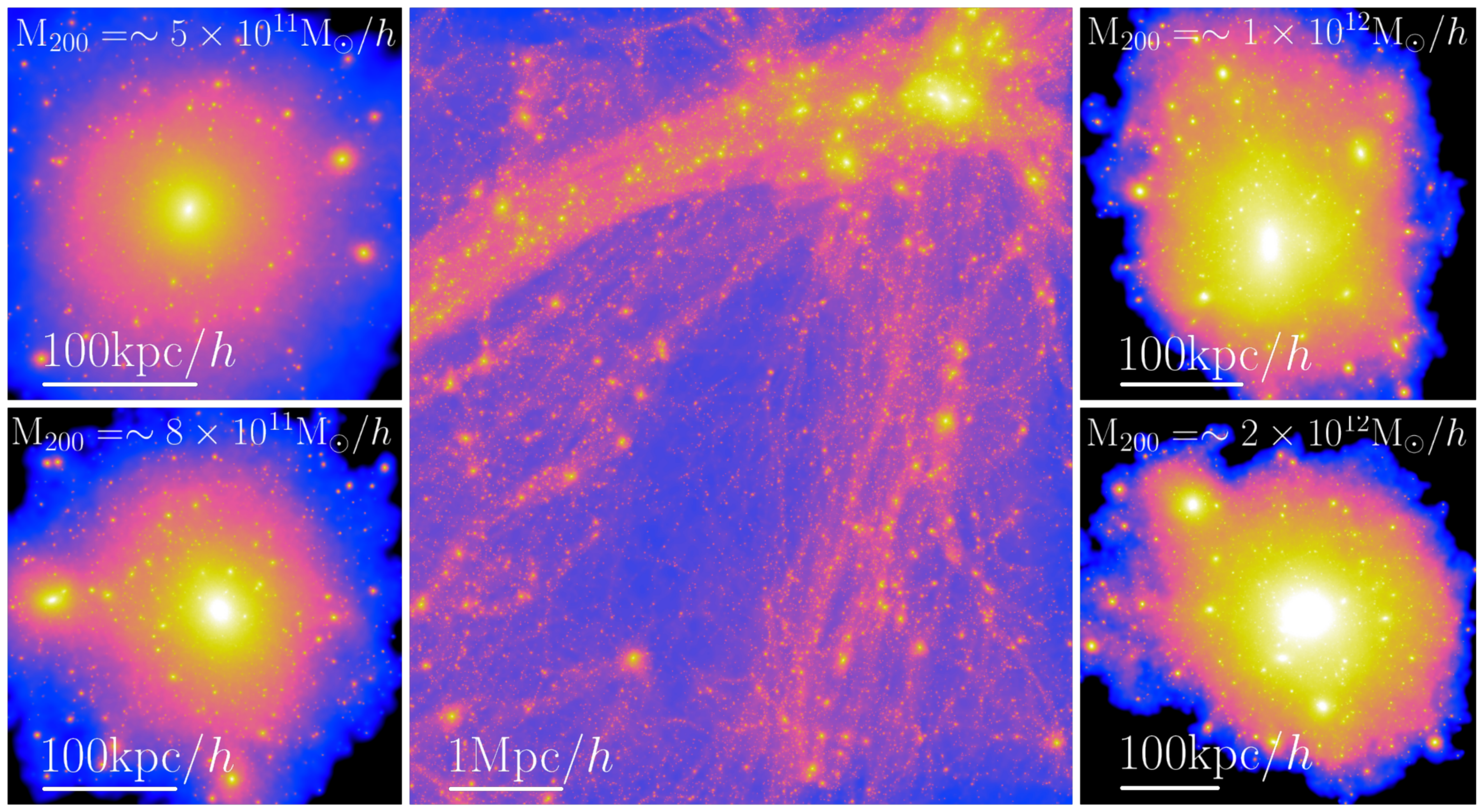}
   \caption{A selection of non-linear structures from the \coco{} simulation. The left- and right-hand 
panels show the projected DM density for four MW-mass FOF haloes. The central panel 
gives the projected DM density (in a $1.5\hmpc$ thick slice) in a region that clearly illustrates the rich
hierarchy of non-linear structures resolved in \coco{}, from massive haloes 
(upper-right corner) to cosmic web filaments and voids.}
\label{fig:coco_dens_2}
\end{figure*}

The \coco{} and \scolor{} simulations were both run with \gadgetIII{} Tree-PM N-body code, which is an updated 
version of the publicly available \gadgetII{} code \citep{GADGET2}. \gadgetIII{} is a hybrid code in which the 
long-range forces are computed using a particle-mesh method, while short-range forces are obtained 
by using a hierarchical oct-tree algorithm. This particular heterogeneous architecture allows for a relatively 
easy follow-up of nested grids placed with increasing accuracy around the high-resolution region. This results in
a proper long-range force accuracy throughout the box, while focusing most of the computational effort inside
the high-resolution region of interest. For both simulations DM particle positions and velocities were saved 
in 160 equally spaced in $log(z+1)$ snapshots. Table \ref{tab:sims_details} summarizes some details 
of the \coco{} and \scolor{} simulations and compares these with some other widely used cosmological simulations.

\subsection{Halo and subhalo finding}
\label{subsec:halofinding}
We identified DM haloes and subhaloes using the \subfind{} algorithm \citep{SUBFIND}. Due to the large number of particles
and high clustering level of the \coco{} simulation, the standard version of \subfind{} would have required a vast amount
of computer memory and CPU time. To overcome this problem, we have used an updated version 
of the algorithm that has been especially optimised for parallel computing and big data. While these changes significantly 
decrease the required computational resources, they do not affect the final output of the method, with the 
new \subfind{} version producing the same halo and subhalo catalogues as the older version. 

\subfind{} starts by identifying DM haloes using the friends-of-friends (FOF) algorithm \citep{Davis1985}, for which we 
used a linking length $b=0.2$ times the mean inter particle separation. If any resulting FOF group has one or more
low resolution particles, we exclude it from further analysis since the internal properties of such objects might have
been affected by unrealistic two-body scattering and self gravity.
All pristine FOF groups with at least 20 particles were kept for further analysis. At $z=0$ we found more 
than $1.2\times 10^7$ ($5.18\times 10^6$) FOF groups in the \coco{} (\scolor{}) run, with the peak 
value was found at $z=3$ and consisted of $1.63\times10^7$($6.19\times10^6$) groups. 
\subfind{} further analyses each FOF group to find gravitationally self-bound DM
subhaloes (i.e. substructures within the FOF groups). Subhalo candidates are first identified by looking for 
overdense regions inside the FOF groups that are further pruned by checking which ones are gravitationally
self-bounded objects. This results in a catalogue of self-bounded structures containing at least 20 particles.
For each subhalo we also compute and store a number of additional properties. 
This consists of peak circular velocity, $V_{max}$, and the physical radius, $R_{max}$, at which this peak 
is attained, half-mass radius, spin (angular momentum), position (corresponding to the minimum of 
gravitational potential) and bulk velocity of the subhalo.

Each subhalo is characterised by a well defined mass, $M_{sub}$, and radius, $r_{sub}$. The former is 
given by the mass contained in all the particles that pertain to the subhalo. The latter is approximately
the subhalo proper tidal radius (see Figure~15 of \cite{Aquarius}).
The FOF groups are characterised in terms of their FOF mass, $M_{FOF}$, as well as of their $M_{200}$ mass.
The first, similarly to subhaloes, is given by the mass contained in all the particles associated to a given FOF group.
In contrast, $M_{200}$ is the mass contained in a sphere of radius $r_{200}$ centred on the FOF group, such that
the average overdensity inside the sphere is $200$ times the critical closure density, $\rho_c$. 
We refer the reader to \citet{Sawala2013} for a comparison of systematic differences between the two as well as other halo mass definitions. 

To compute the radial profiles of haloes, we follow a prescription similar to that 
employed by \citet{Power2003} and \citet{AHF}. Namely, we identify the centre of mass of FOF groups using an iterative
procedure, by computing the centre of mass inside smaller and smaller spheres, with each such sphere 
centred on the centre of mass found in the previous iteration step. The centre of each FOF group is 
used to grow logarithmically spaced spherical shell bins up to $r_{200}$. Figure~\ref{fig:coco_dens_2}
illustrates the level of detail to which we resolve MW-mass FOF haloes.

\subsection{Merger Trees}
\label{subsec:merger_trees}

In CDM cosmologies the first objects to from are DM clumps
(haloes) with Jeans mass of the order of Earth mass $\sim 10^{-6}\textrm{M}_{\odot}$ \citep[see \eg][]{Green2004}.
Due to numerical limitations, such small density perturbations are not resolved in our simulations, 
hence the first objects to from in \coco{} have masses of $\sim 10^{6}\Msun$, some 12 orders of 
magnitude larger. However, it is well established \citep[\eg][]{Mtree_Lacey1993,Merging_Kauffmann1993,Mtree_Roukema1997} 
that in hierarchical cosmologies characterised 
by a nearly scale-free Harrison-Zeldovitch like \citep{HarrisonSpec,ZeldovichSpec} initial power spectrum, such as $\lcdm$, 
larger objects forms by consecutive merging of smaller ones. Successive populations of haloes grow 
from mergers of earlier populations accompanied by accretion of some smooth mass component 
\citep[see \eg][]{Wechsler2002}.
In order to trace the temporal evolution of haloes we constructed DM haloes 
{\it merger trees} \citep[for more details see][]{MTrees1}. For this, we employed a recently updated algorithm 
that has been developed for use with the semi-analytic galaxy formation code \verb#GALFORM# \citep{SAMSCole2000}.
The method we used is described in detail in \cite{Mtrees3} and is 
an upgrade over the earlier version of \cite{Mtrees2}. The essential part of the algorithm consists of 
unique linking between subhaloes from two consecutive snapshots. This allows for a construction of 
very precise merger trees at the subhalo level. We have applied this algorithm to the \coco{} simulation, 
resulting in approximately $1.312\times 10^9$ unique subhaloes contained in the merger trees.

\section{Dark Matter haloes}
\label{sec:haloes}

In this section we focus on a few key aspects of DM haloes: their abundance as a 
function of mass, their internal structure and their formation histories. 
Understanding the basic properties of DM haloes is a key ingredient of any successful galaxy formation theory,
since galaxies are formed and evolve inside their host haloes.  
Furthermore, understanding the link between the properties of DM haloes and the luminous galaxies that reside within them
is crucial for designing and conducting astrophysical tests of the $\lcdm$ paradigm.

\subsection{Mass function}
\label{subsec:mass_func}
\begin{figure}
  \includegraphics[width=85mm]{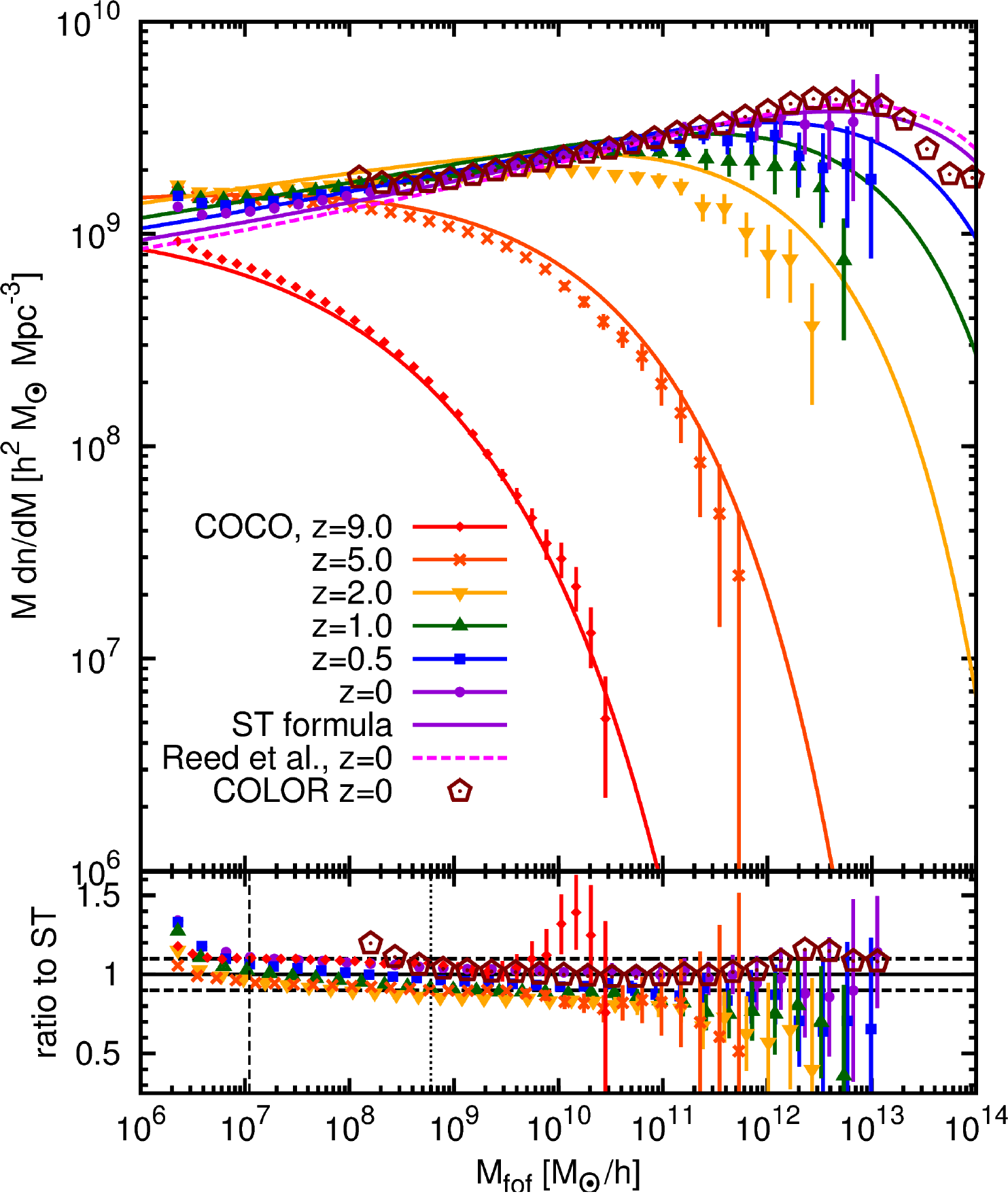}
   \caption{{\it Upper panel:} FOF mass functions for the \coco{} and the \scolor{} simulations, 
   on which we superimpose the Sheth-Tormen prediction (solid lines). The \coco{} results and the ST predictions 
 are plotted for a wide range of redshifts, from $z=9$ (red diamonds) to $z=0$ (purple circles). For comparison 
 we also plot the $z=0$ results from the \scolor{} run (open hexagons) and the Reed {\it et al.} prediction
(dashed line). The vertical bars indicate Poisson errors.
{\it Bottom panel:} The \coco{} and \scolor{} mass functions normalized by the ST prediction
at that redshift. The horizontal dashed lines mark a $10\%$ difference level. The vertical dashed (dotted) line illustrate the \coco{} (\scolor{}) 
halo mass resolution limit.
}
\label{fig:coco_MF}
\end{figure}

Accurate theoretical predictions for halo mass functions are needed for a number of reasons. For example, 
they are a primary input for modelling galaxy formation, whether it be physically motivated semi-analytical 
models \citep[\eg][]{SAMSCole1994,SAMSCole2000} or statistical-based approaches like abundance matching 
\citep[\eg][]{YangAM2003,GuoAM2010}.
The abundance of haloes across cosmic epochs was studied since the
early work of \cite{Press-Schechter}, with predictions resulting from the extended excursion 
set models based on ellipsoidal collapse \citep{MFShethTormen} or 
motivated by N-body simulations \citep[\eg][]{MFJenkins2001, Warren2006, MFReed2007}.
Such models were thoughtfully tested in computer simulations, but only in a limited mass range 
$M_{\rm FOF}>\textrm{a few}\times10^8\Msun$. Due to our unique \coco{} and \scolor{} simulations, we are able
to investigate the abundance of FOF haloes down to lower masses and over a wide range, spanning
8 decades in halo mass. 

In Figure~\ref{fig:coco_MF} we compare the present day \coco{} and \scolor{} halo mass function with
the Sheth\&Tormen (ST) prediction \citep{MFShethTormen} and with the improvement suggested
by \citet[][hereafter R07]{MFReed2007}, which was tuned using results of N-body simulations.
The R07 prediction includes the dependence of the halo mass function on the effective power spectral 
slope ($n_{eff}$) at the scale of the halo radius.
We find good agreement between the present day \coco{} and \scolor{} mass functions all the way from 
the resolution limit of the \scolor{} simulation, $M_{\rm FOF}\sim6\times10^8\Msun$, up to the most massive objects found in 
the \coco{} volume, $M_{\rm FOF}\sim10^{13}\Msun$. As the resolution limit of FOF haloes we adopt a minimum 
threshold of 100 particles. Note that this is different from the resolution limit of converged internal (sub)halo 
properties (\eg~$V_{max}$) that we derive in Appendix \ref{sec:appendix}.
This assures as that the specific choice of the \coco{} 
region (see \S\ref{sec:simulation}) did not introduce any significant halo abundance bias (scarcity or excess) 
for the range of halo masses that we are interested in. We also see a good agreement with the ST and R07 models at $z=0$, though
both \coco{} and \scolor{} predict slightly more low mass haloes. This discrepancy is rather small, for example at a halo 
mass of $10^7\Msun$, where the difference is the largest, ST predicts a $\sim11\%$ lower halo abundance, 
while the R07 result is $\sim18\%$ lower. These differences are unlikely to be caused by numerical effects. 
First, while it is well known that the FOF algorithm tends to overpredict the abundance of poorly resolved haloes, 
this effect is significant only for objects with fewer than 100 particles \citep{Warren2006}. 

Figure~\ref{fig:coco_MF} also shows the time evolution of the \coco{} halo mass function from redshift $z=9$ till the present day. 
The result beautifully reflects the well known hierarchical character of DM halo build-up, with smaller haloes
forming first, which in turn merge into bigger and bigger objects. For comparison for each \coco{} redshift data we also plot
the ST prediction line. It is clear that the ST prognosis fails to match the \coco{} data for $0.5\leq z\leq 5$ as it significantly
overpredicts the abundance of objects. Similar results were also found by \citet{Bolshoi}.
Interestingly this discrepancy is largest for the intermediate redshifts of $1\simlt z\simlt2$, 
while at $z=9$ the ST forecast is again in a good agreement with our data. 
The epoch at which we observe the biggest discrepancy
between the \coco{} data and the ST predictions also happens to be the epoch at which the halo merger rates are 
the highest \citep[see \eg][]{Hopkins2010},
hence the mass function of collapsed objects experience the most dynamical evolution, which in turn is 
reflected in the failure of the ST forecast.
\begin{figure}
  \includegraphics[width=85mm]{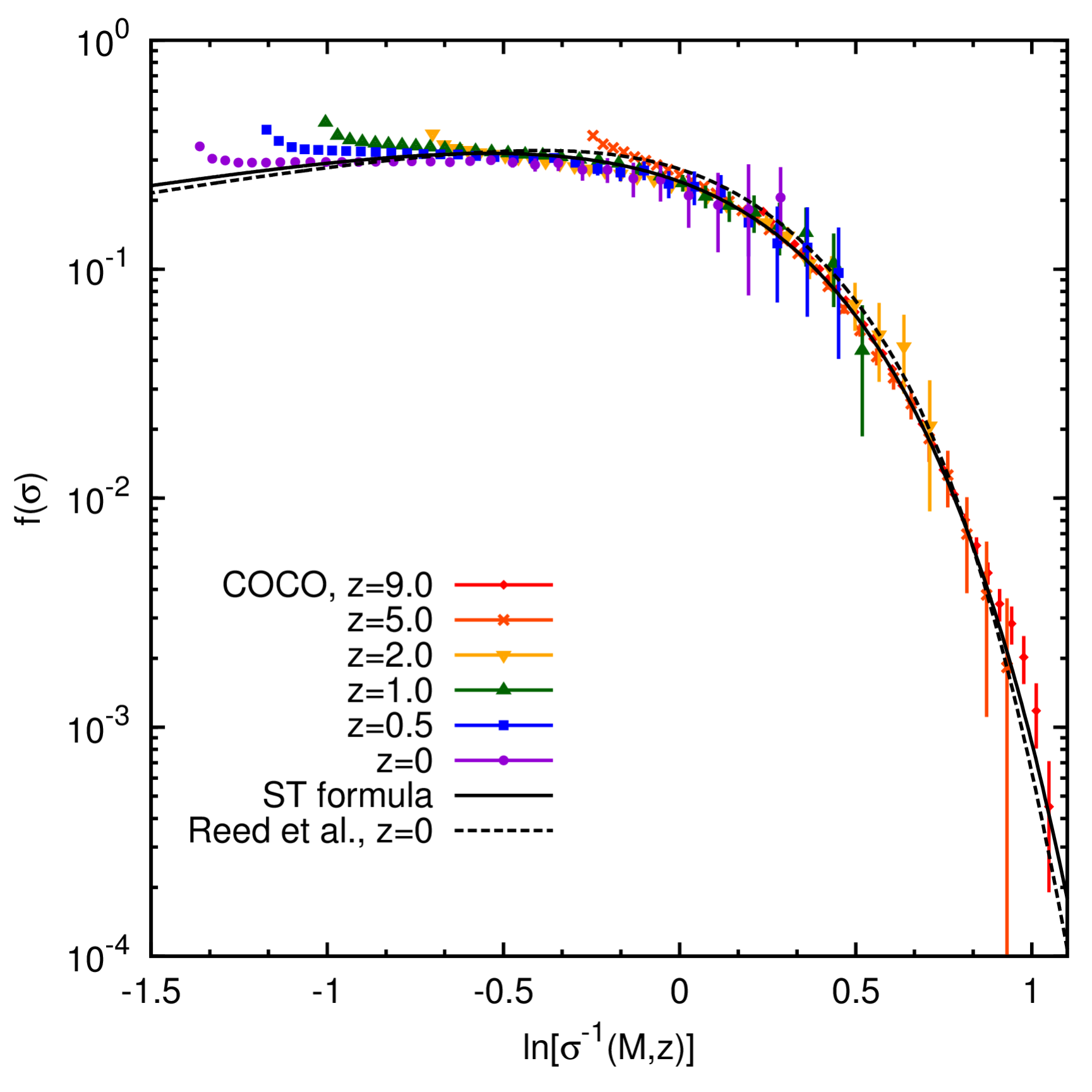}
   \caption{FOF halo multiplicity function, $f(\sigma)$, as a function of peak mass variance, $\sigma(M,z)$, for a range of redshifts.
    The vertical bars indicate Poisson errors.
}
\label{fig:coco_multiplicty}
\end{figure}

To better compare with analytical models for the abundance of haloes,
it is more convenient to express the halo mass, $M$, in terms of 
the variable $\ln[ \sigma^{-1}(M,z) ]$, where $\sigma(M,z)$ gives the peak mass variance at scale $M$ 
and redshift $z$. This quantity is defined as
\be
\sigma^2(M,z) = {1\over 2\pi^2}\int P(k,z)W^2(k,M)k^2\textrm{d}k\,,
\label{eqn:smoothed_variance}
\ee
where $P(k,z)$ is the power spectrum of linear density fluctuations extrapolated to redshift $z$ and $W(k,M)$ is 
the Fourier transform of the top-hat window corresponding to the radius enclosing mass $M$ 
at the mean density of the universe. Now, the halo mass function can be written as
\be
M{\textrm{d}n(M,z)\over\textrm{d}M}=\overline\rho(z){\textrm{d}\ln\sigma^{-1}\over\textrm{d}M}f(\sigma)\,,
\label{eqn:universal_MF}
\ee
where $\overline{\rho}(z)$ is the mean mass density of the universe at redshift $z$ and
$f(\sigma)$ denotes the halo multiplicity function. This latter quantity, $f(\sigma)$, takes a universal form
that is independent of redshift \citep[for more details, see][]{MFJenkins2001,MFReed2007,TinkerMF2008,MXXLAngulo2012}.
In Figure~\ref{fig:coco_multiplicty} we plot the multiplicity function of FOF haloes as a function of $\ln[ \sigma^{-1}(M,z) ]$
at various redshifts. Independent of redshift, the data points follow, 
to a good approximation, the universal shape as predicted by both ST and Reed \etal{} formulas.

\subsection{Density profiles and the mass-concentration relation}
\label{subsec:density}

Spherically averaged radial density profiles are one of the simplest yet robust characterisations 
of the internal structure of DM haloes. It is well established that for hierarchical cosmologies like CDM, 
the radial density profile of relaxed DM haloes 
is to a good approximation self-similar and can be mapped 
by a simple broken power-law formula, the NFW profile \citep{NFW1, NFW2}:
\be
\label{eqn:nfw}
{\rho(r)\over\rho_{crit}}={\delta_c\over(r/r_s)(1+r/r_s)^2}\,.
\ee
It characterises the profile of any halo by 
two parameters: a scale radius, $r_s$, and a characteristic overdensity, $\delta_c$. 
Instead of working with the $r_s$ and $\delta_c$ parameters, it is customary to define
the halo concentration, $c_{200}$, as:
\be
\label{eqn:concentration}
c_{200} = {r_{200}\over r_s}\,,
\ee
with $r_{200}$ the virial radius of the halo defined in \S\ref{subsec:halofinding}.
Using this parametrisation, the NFW profile effectively becomes a one parameter fit,
since the characteristic overdensity can be expressed as:
\be
\label{eqn:delta_c-c}
\delta_c={200\over 3}{c_{200}^3\over \ln(1+c_{200})-c_{200}/(1+c_{200})}\,.
\ee
The density profile of a halo is also probed by the shape of the circular velocity curve, which is:
\be
\label{eqn:circ_vel}
V_{c}(r) = \sqrt{{\textrm{G}M(<r)\over r}}\,,
\ee
where $M(<r)$ is the mass contained inside a sphere of radius $r$ centred at the halo centre. For a perfectly 
spherical halo, the circular velocity, $V_c(r)$, is exactly equal to the circular orbital velocity 
at distance $r$. For well resolved and relaxed haloes, the circular velocity takes only one maximum 
value, $V_{max}$, that is attained at radial distance, $R_{max}$. Similar to the virial mass, we can define 
the virial circular velocity $V_{200}=(\textrm{G}M_{200}/r_{200})^{1/2}$.
The circular velocity, $V_c(r)$, bears effectively the same information as the halo density profile, $\rho(r)$, 
but is much less prone to noise because of the integral nature of the former.

The NFW scale radius, $r_s$, gives the radial position at which the $r^2\rho(r)$ curve attains
its maximum, which sometimes is also denoted by $r_{-2}\equiv r_s$. For the majority of DM haloes, 
the peak of the $r^2\rho(r)$ curve is relatively broad. This means that, for haloes resolved
with a relatively small number of particles, the exact location $r_{-2}$ of the peak is 
uncertain due to the presence of noise. This is reflected in the susceptibility of the NFW fit to the radial 
range used for fitting the profile of haloes resolved with fewer than a few thousand 
particles \citep{Navarro2004,Prada2006,CMGao2008,Ludlow2010}.
This is especially prominent when profiles of many similar mass haloes are stacked 
to remove halo-to-halo variation due to the presence of substructures. This behaviour indicates 
that fitting NFW profiles to haloes resolved with a relatively small
number of particles is biased and gives rise to an artificial correlation between concentration and halo mass. 
As a solution, \cite{Navarro2004} proposed the use of a more flexible parametrization,
that would account for the differences between NFW and stacked universal halo profiles. 
The improved three-parameter fitting formula takes
a form, in which the logarithmic slope of density assumes a single power law:
\be
\label{eqn:Einastio1}
{\dd\log \rho(r)\over\dd\log r}=-2\left({r\over r_{-2}}\right)^{\alpha}\,.
\ee
This induces radial density profile of the form:
\be
\label{eqn:Einastio2}
\ln(\rho(r)/\rho_{-2})= -(2/\alpha)[(r/r_{-2})^{\alpha}-1]\,,
\ee
where $\rho_{-2}$ is the density at $r_{-2}$. The additional parameter $\alpha$ 
is called the shape parameter and, for CDM haloes, it typically takes values in the range $0.1 - 0.3$. 
This power-law density profile was first introduced by \cite{Einasto1965}
to model the density distribution of the stellar halo of our own Galaxy. To distinguish this 
density fitting function from the NFW profile we will refer to it as the Einasto profile. 
The Einasto profile can be characterised in terms of a concentration parameter that is given
by eqn.~(\ref{eqn:concentration}) with $r_s$ replaced by $r_{-2}$.
Both the Einasto $r_{-2}$ parameter as well as the NFW scale radius, $r_s$, correspond
to the scale at which the logarithmic slope of the density profile 
attains the 'isothermal' value of -2. In addition, for $\alpha\simeq0.2$ 
the Einasto profile approximates fairly well the NFW profile in the fit range.

In this work we are interested in a statistical description of DM haloes concentrations, 
with emphasis on the relation between concentration and halo mass,
its variance and its redshift evolution. To obtain robust measurements, we fit 
both the NFW and Einasto profiles to all the haloes with at least $N^{min}_p=5000$ particles,
which for \coco{} corresponds to a minimum halo mass, $M^{min}_{200}=5.7\times10^{8}\Msun$. 
We discuss further down why we picked this particular limiting value.
The fitting procedure finds the parameter values that minimize the merit function
\be
\label{eqn:fit_merit}
\sigma^2(\vec{\Xi})=N_{bins}^{-1}\sum_{i=1}^{N_{bins}}[\ln\rho_{i}-\ln\rho_{fit}(\vec{\Xi})]^2\,,
\ee
where the vector of fit parameters $\vec{\Xi}=(r_s,\delta_c)$ and $(r_{-2},\rho_{-2},\alpha)$ for
the NFW and Einasto fits, respectively. The number of radial bins, $N_{bins}$, is equally spaced
in $\log(r)$ and is selected adaptively depending on the number of particles, $N_p$, contained
in the halo as:
\be
\label{eqn:nbins}
N_{bins} = 2\textrm{ceiling}(6.2\log{N_p}-3.5)\,.
\ee
This gives $N_{bins}=92$ for our most massive halo and $N_{bins}=40$ bins for a halo with 5000 particles.
Since the bins are equally spaced in $log(r)$, the inner bins
contain significantly fewer particles than the mid-range and outer bins. Hence, as we move towards the halo centre,
the radial bins become more affected by sampling noise and two body scattering effects. 
This has been studied by \cite{Power2003}, who 
found that there is a minimum radius below which one cannot trust the radial density profile
of haloes extracted from N-body simulations. This convergence radius
is given by the inner most bin that fulfils the \cite{Power2003} criterion (see eqn.~(20) in their paper). 
We exclude from the fitting all radial bins that are below the convergence radius for a given halo.
After applying this convergence criterion, a halo resolved with $N^{min}_{p}=5000$ particles is
left on average with 20 radial bins. Thus, the minimum halo mass for which we perform a fit
is given by the mass for which more than a half of the radial bins pass the convergence test.

The individual density profile of haloes resolved with fewer than $N^{min}_{p}=5000$ particles is very sensitive to
the intrinsic numerical noise. However, there is still plentiful of information that can be 
extracted from haloes with fewer than $N^{min}_{p}$ particles. By stacking many such haloes, the noise of
the density profile is significantly reduced. Finding the position of the maximum of the $r^2\rho(r)$ 
curve for a such stacked profile gives the $r_s$ and $r_{-2}$ parameters for the NFW and Einasto profiles, respectively.
We apply this method to get median concentration of stack profiles for haloes with 
$800\leq N_{p}\leq 5000$. Using this approach, we estimate halo concentrations in \coco{} down to a halo mass
of $M_{200}= 9\times10^{7}\Msun$. 

The NFW and Einasto profiles represent a good description of the radial density profiles 
for virialised DM haloes, which are in equilibrium. 
Haloes that experienced recent mergers or close encounters can be far away from the state 
of equilibrium and, thus, their density profiles are usually not
very well described by neither NFW nor Einasto profiles. As a consequence, the concentration 
parameter derived from fitting unrelaxed haloes is ill defined
and at best biased low \citep[see \eg][]{Neto2007, CMGao2008, Ludlow2010}. To overcome this 
problem, we remove non-virialised haloes, i.e. objects that do not satisfy the following three
criteria \citep{Neto2007}:
(i) the fraction of halo mass contained in its resolved substructure is $f_{sub}<0.1$, 
(ii) the displacement between the centre of mass and the minimum of the gravitational potential
cannot exceed $7\%$ of halo's virial radius, $r_{200}$, 
and (iii) we require that the adjusted virial ratio, 
$\nu\equiv(2T-E_s)/|U|$, is $\nu<1.35$. 
Here $T$ and $U$ are the halo's total kinetic and potential energy, respectively. To account 
for the fact that real haloes are not isolated objects, we include the Chandrasekhar pressure 
term, $E_s$, which quantifies the degree to which a given halo interacts with its surroundings.
See \cite{Shaw2005} and their eqn.~(6) for the definition and method used to estimate the pressure term,
and also see \cite{Power2012} for a more detailed discussion about the virial ratio of haloes. 
\begin{figure*}
  \includegraphics[width=175mm]{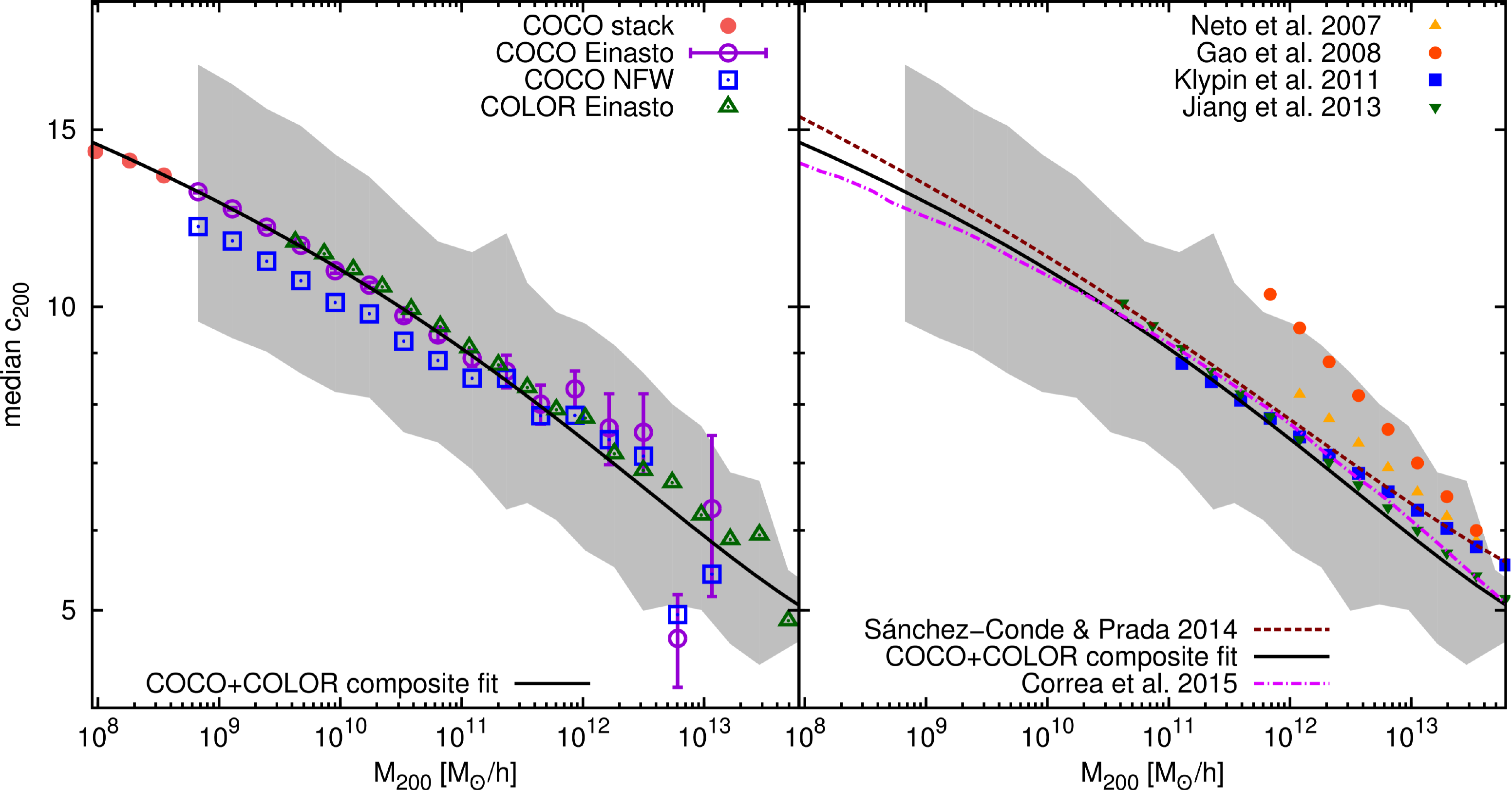}
   \caption{{\it Left-hand panel:} Concentration-mass relation for relaxed haloes.
    The open symbols give the median concentration obtained from fitting NFW and Einasto profiles. 
    The best-fitting line was obtained from the combined Einasto \coco{}+\scolor{} data, with the \coco{} concentrations (circles) extended
    by the \scolor{} results (triangles) for halo masses $M_{200}>5\times 10^{11}\Msun$. 
    The filled circles show the average concentrations obtained from stacking profiles for the three
   lowest mass bins. For comparison we also show the data obtain by fitting the NFW profile (open boxes).
  {\it Right-hand panel:} Comparison of various $c_{200}-M_{200}$ fits from the literature (lines with symbols and the dashed line)
  with our best-fitting line (solid line). In both panels, the shaded region shows the 16th to 84th percentile scatter
   of the \coco{}+\scolor{} data.
   }
\label{fig:coco_c-M}
\end{figure*}

The left-hand panel of Figure~\ref{fig:coco_c-M} shows the median halo concentration
as a function of halo mass for our two simulations, \coco{} and \scolor{}.
We find a very good agreement between \coco{} and 
\scolor{} results for all halo masses up to
$M_{200}\sim3.5\times10^{12}\Msun$. Above this mass, due to scarcity of the massive haloes, 
the \coco{} results are dominated by halo-to-halo scattering.
This is clearly seen from the increasing size of the error bars that show the bootstrap errors of the median. 
The good agreement between the two simulations suggests that we can supplement 
the \coco{} data at the high-mass end by adding the objects from the \scolor{} simulation.
We construct such a joint sample and use it for obtaining our best-fit for the median $c_{200}-M_{200}$ relation as discussed later.
Comparing the results for the NFW and Einasto fits, we find clear differences between the two.
Below a halo mass of $M_{200}\lesssim10^{11}\Msun$, the difference takes the form of a
systematic shift, with the slope of the $c_{200}-M_{200}$ relation being similar for both profile fits.

In the right-hand panel of Figure~\ref{fig:coco_c-M} we also show various fits to the $c_{200}(M_{200})$ relation
with the goal of comparing the accuracy of these literature fits with the results of N-body simulations.
From the set of single power-law fits, the one that best matches the data is the fit proposed by 
\citet{Mtrees3}. This is in very good agreement (better than $4\%$) with both \coco{} and \scolor{} data down to a halo
mass of $\sim2\times 10^{10}\Msun$. It seems that below that mass our data indicate slight flattening 
of this relation, hence change of the slope. Also the fit of \citet{Bolshoi} is in a reasonably good agreement
with our data, except for the most massive objects ($\sim 5\times 10^{13}\Msun$) where it predicts 
a median concentration that is higher from the value of $84\%$ of our haloes at that mass.
The Klypin et al. results were found for halo masses and boundaries
defined by a spherically averaged overdensity 
$\Delta_{vir}=360\times\Omega_{m,0}\times\rho_{c}$. This definition roughly corresponds to
$\Delta_{100}$ in our nomenclature, hence, to allow for a comparison of their fit 
with our data, we have rescaled their $c_{vir}-M_{vir}$ relation to
appropriate equivalent of $c_{200}-M_{200}$. However, this procedure ideally should 
be performed for each halo separately at the particle level, so our
rescaling here can only be treated as an approximation. The performance of the \cite{Neto2007} fit 
is also reasonably good down to $M_{200}\sim10^{12}\Msun$.
The fit of \cite{CMGao2008} agrees with our data only for the most massive bins and it clearly predicts a different 
slope of the concentration-mass relation. This unavoidably 
leads to a significant overestimation of the median halo concentration
by their fit for haloes below a mass of $\sim2\times10^{12}\Msun$. There are two 
possible sources driving this discrepancy. 
First, the Gao et al. fit is based on the \MI{} which uses significantly different 
values of the cosmological parameters. This difference is most notable for the $\sigma_8$ parameter that 
is $10\%$ higher in the \MI{} than in our simulations.
The different cosmology is probably the main reason for the observed differences since variations 
in both $\Omega_m$ and $\sigma_8$ have a large impact on halo concentrations \citep{LudlowPMAH2013,Dutton2014,Diemer2015}.
Secondly, the Gao et al.
relation is obtained by fitting a relatively limited
halo mass range given by $5\times10^{12}\leq M_{200}/(\Msun)\leq10^{15}$. 
Since the concentration-mass relation is not a simple power law, fitting a single power
law provides a relation that holds only for that mass range \citep{LudlowMC2014,Sanchez-Conde2014, Prada2012}.
\begin{table*}
\begin{center}
\caption{Our best-fitting parameters for the $c_{200}-M_{200}$ relation for the functional from of the Eqn.(\ref{eqn:c_M_fit}).
}
\label{tab:cMfit_param}
\begin{tabular}{@{}cccccc}
\hline\hline
$c_0$ & $c_1$ & $c_2$ & $c_3$ & $c_4$ & $c_5$ \\
\hline
34.988 & -1.9841 & $8.039\times10^{-2}$ & $-1.777\times 10^{-3}$ & $-1.4557\times 10^{-5}$ & $7.34152\times 10^{-7}$\\
\hline
\end{tabular}
\end{center}
\end{table*}

To emphasise this last point, we also checked the performance of the multi 
power-law median $c_{200}(M_{200})$ model of \citet[][hereafter SC14]{Sanchez-Conde2014}, based on a functional 
form proposed by \cite{Lavalle2008}:
\be
\label{eqn:c_M_fit}
\langle c_{200}(M_{200},z=0)\rangle_{med}=\sum_{i=0}^5 c_i\times[\ln(M_{200}/\Msun)]^i\,.
\ee
This multi-component fit is claimed to be a much better match for the median halo 
concentration-mass relation due to flattening of this relation at small halo masses.
Not surprisingly, the SC14 fit shows a very good agreement with the N-body results
to an accuracy better than $10\%$. However, the SC14
fit systematically overpredicts the median concentration for our haloes with 
$M_{200}<10^{11}\Msun$. Especially for the first four to five least massive bins,
we can notice the difference of varying slope of the concentration-mass relation 
between the SC14 fit and our \coco{}+\scolor{} sample. To better quantify 
this discrepancy, we have fitted eqn.~(\ref{eqn:c_M_fit}) to a combination of \coco{}+\scolor{} data.
This '\coco{}+\scolor{} composite' set was obtained by augmenting the \coco{} data for halo masses above $5\times 10^{11}\Msun$
with the \scolor{} data and further supplementing it at the low-mass end with the data obtained from the profile stacking.
Our best fitting parameters are presented in the Table~\ref{tab:cMfit_param}
and the corresponding fit is marked as a solid black line in Fig.~\ref{fig:coco_c-M}. 
To fix the asymptotic freedom of the fit at 
the low-mass tail, we have used the data points from \cite{Diemand2005,Ishiyama2013} and \cite{Anderhalden2013}. 
For haloes with $1\lesssim M_{200}/(\Msun)\lesssim 10^{6}$ our fitted relation predicts 
concentrations that are
$\sim10\%$ lower than the SC14 fit, while for even smaller haloes this tendency 
flips and our fit predicts concentrations that are systematically higher. However,
such a difference has only small effects on the predicted boost factors for the radiation flux of DM annihilation.
For completeness, we also compare the median \coco{} $c(M_{200})$ relation with the model 
of \cite{CorreaIII2015} (hereafter C15) which is based on the mass accretion 
history of haloes \citep[see also][]{CorreaI2015,CorreaII2015,LudlowMC2014}.
The prediction of the C15 model for a WMAP7 cosmology is
shown in the right-hand panel of Fig.~\ref{fig:coco_c-M} as the dashed-dotted line.
This model agrees to better than $5\%$ with our data for haloes more massive than $\sim10^9\Msun$. At
lower masses, the model underpredicts the concentration, which most likely reflects the fact that 
the C15 model was built for NFW profiles rather than for Einastio profiles.
\begin{figure}
  \includegraphics[width=85mm]{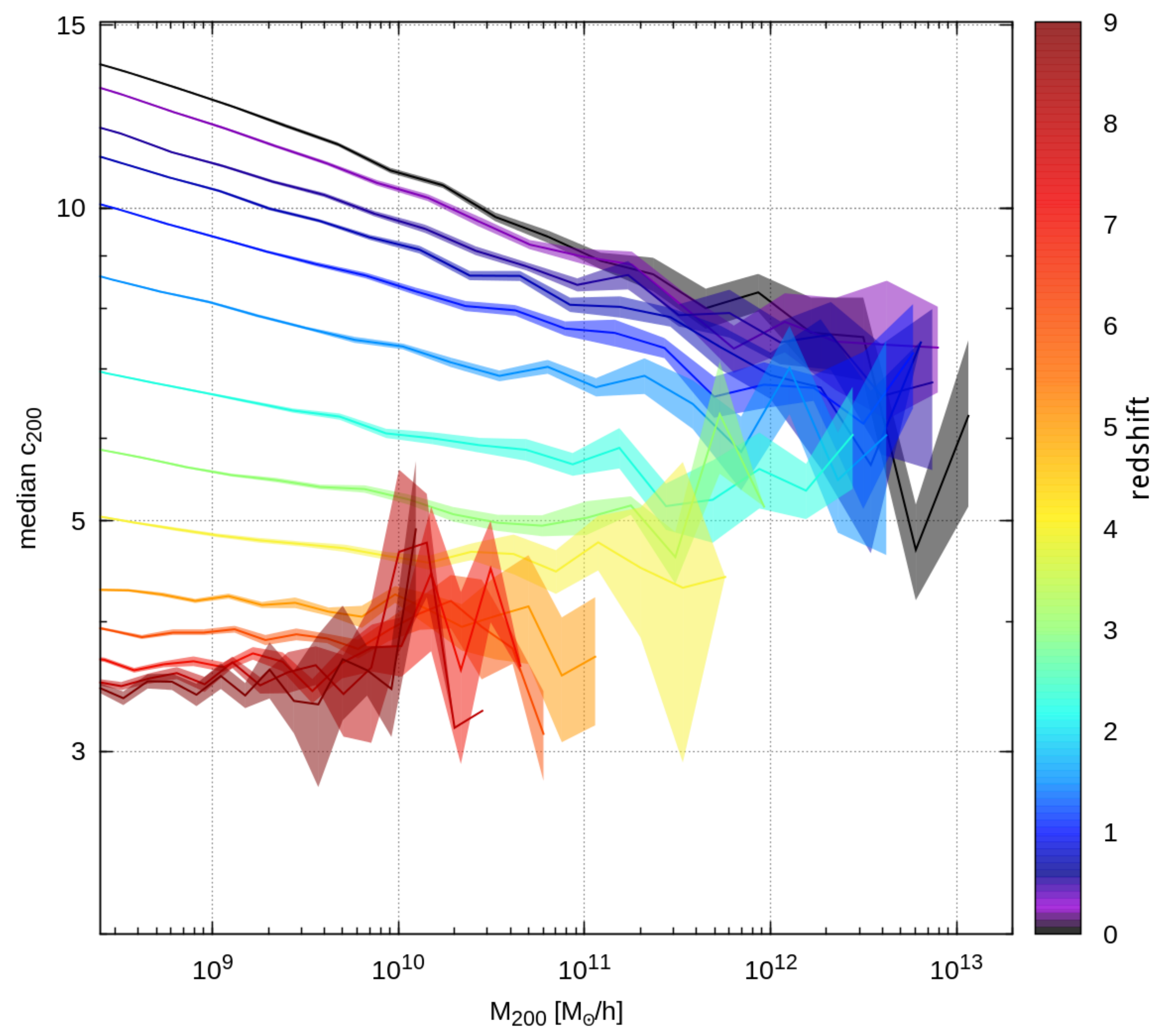}
   \caption{The redshift evolution of the concentration-mass relation for relaxed haloes in the \coco{} volume. 
    We plot the data from $z=9$ (red colour) down to $z=0$ (black). Solid lines mark the median Einasto concentration,
    while the filled regions show the $1\sigma$ uncertainty in the relation.
}
\label{fig:coco_c-M-evolution}
\end{figure}
\begin{figure}
  \includegraphics[width=85mm]{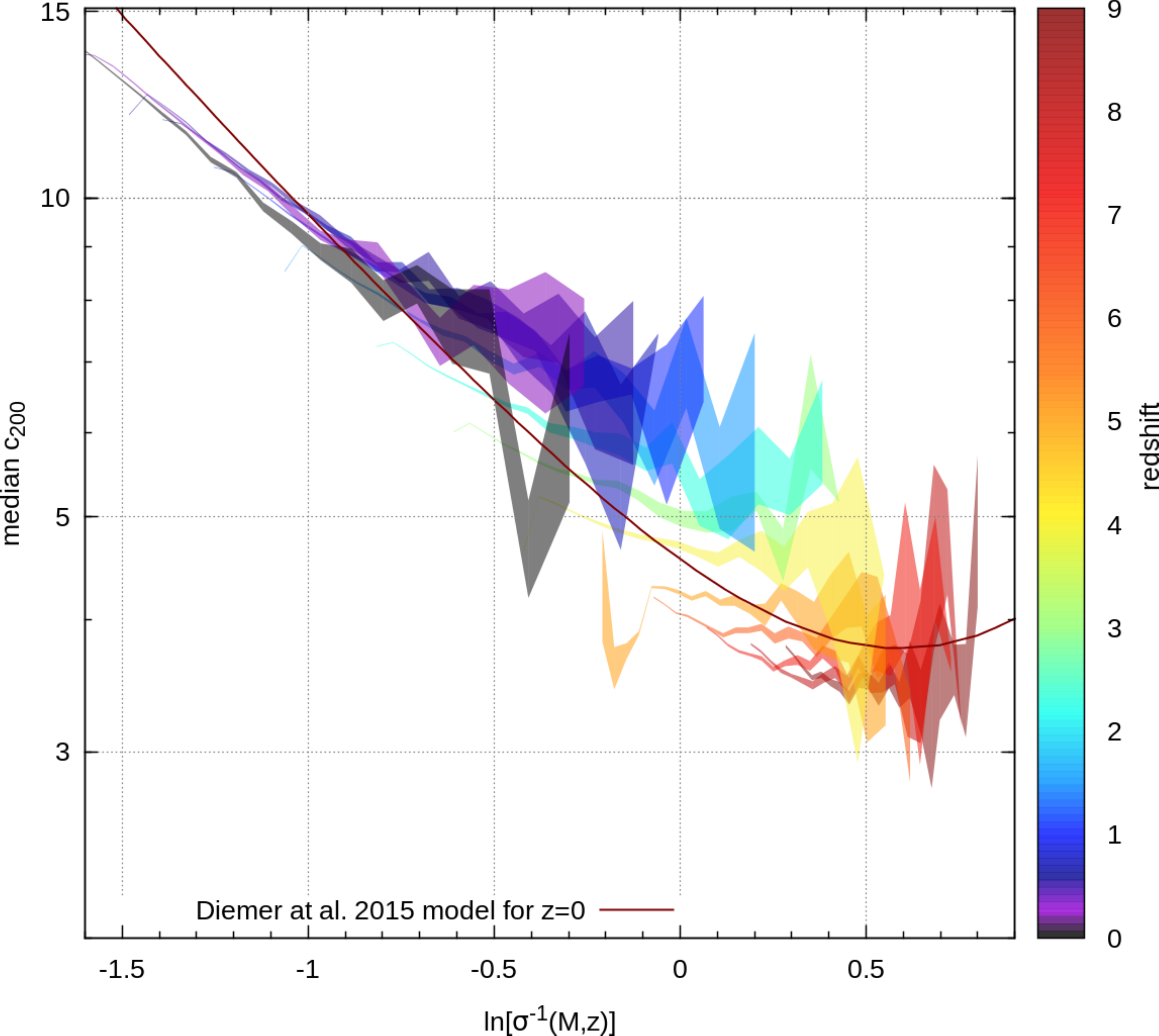}
   \caption{ The halo concentration, $c_{200}$, as a function of peak mass variance, $\sigma(M,z)$, for relaxed \coco{} haloes.
    We show the relation at various redshifts, from $z=9$ (red) down to $z=0$ (black). The black solid line gives the prediction of the \citet{Diemer2015}
    model at $z=0$.
}
\label{fig:coco_c-M-evolution_sigmaM}
\end{figure}

In Figure~\ref{fig:coco_c-M-evolution} we depict the time evolution of the concentration-mass relation.
For $z\gtrsim2$, we find a flattening of the relation at the high mass end, with the flattening moving towards lower masses
for higher redshifts. The same flattening is present also at $z\simeq0$, but we do not see it in the \coco{} data
since the simulation does not resolve very high mass haloes. 
The concentration is related to the characteristic density of a halo, which in turn reflects the mean 
density of the universe at the time when the central part of the halo has collapsed \citep[\eg][]{CMGao2008,LudlowMC2014}.
This naturally leads to the observed flattening of the $c(M,z)$ relation for very rare and massive objects,
since these have assembled only recently and therefore share the same collapse time. 
We also find an evolution in the slope of the relation for lower mass haloes. 
This is in agreement with the well established picture according to which haloes
are build-up hierarchically. Haloes continuously increase their mass with time, so the concentration at fixed
halo mass is determined by different objects for each redshift bin. 
To better understand the time variation of the halo mass-concentration relation, we express the 
halo mass in terms of the mass variance, $\sigma(M,z)$ (see Eqn.~(\ref{eqn:smoothed_variance})). The corresponding
median $c[\sigma^{-1}(M,z)]$ relation is given in Figure~\ref{fig:coco_c-M-evolution_sigmaM} and 
shows that at fixed $\sigma(M)$ values the concentration varies only slowly with time. 
For comparison, we also give the prediction of the \citet{Diemer2015} model to find that 
in the `big-peak' regime (haloes corresponding to rare peaks at a given epoch) our results
are in reasonable agreement with their prediction. At small $\sigma^{-1}(M,z)$ values our data suggest a
less steeper slope. The difference can be accounted for by noting that \citet{Diemer2015} have used a different 
halo finder and that their simulations had much lower mass resolution, hence
they could not probe the 'very small halo' regime, which our simulation resolves.

\subsection{Formation times}
\label{subsec:formation_times}
\begin{figure}
  \includegraphics[width=85mm]{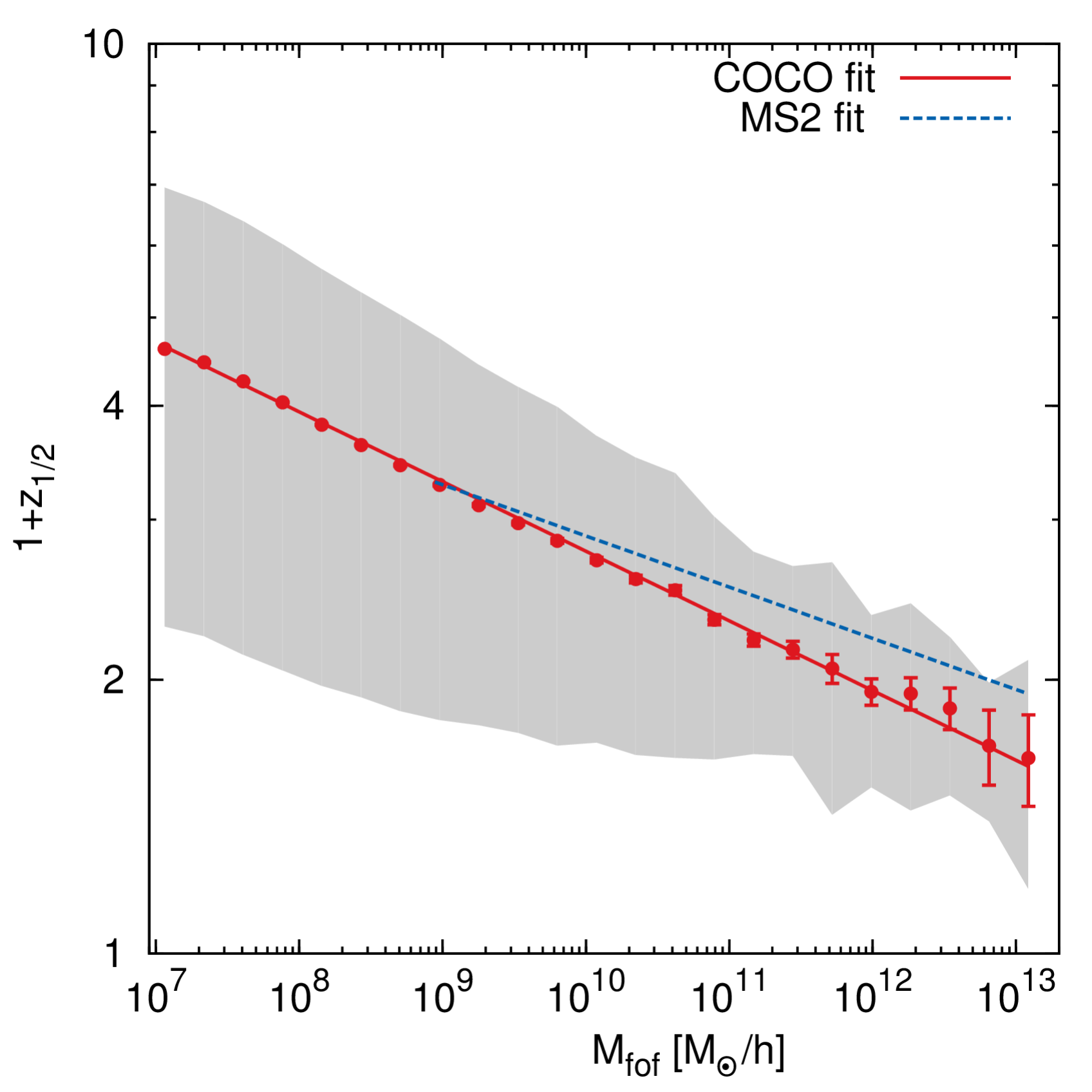}
   \caption{The mean half-mass formation time, $z_{1/2}$, as a function of present day halo mass,$M_{FOF}$. 
   The red points are the simulation data with error bars reflecting the bootstrap error on the mean. 
   The shaded grey region marks the $1\sigma$ scatter around the mean. 
   The solid lines illustrate the fit to the \coco{} (red line) and to the \MII{} (blue line) data.
   }
\label{fig:coco_c-M-zform}
\end{figure}

The highly hierarchical character of structure formation is especially strongly imprinted
in the build-up and mass accretion histories of dark haloes. A significant fraction 
of a halo's mass is assembled via the accretion of other haloes so it is natural 
to expect that more massive objects form later than the low mass ones, as
confirmed by many N-body simulations of the CDM model.
However, the precise form of the halo mass-formation time relation and its intrinsic scatter 
is still a subject of discussion\citep{Mtree_Lacey1993,Wechsler2002}. Determining this relation is relevant for a number of reasons.
Among which, most importantly, is understanding how well this 
relation is correlated with the properties of the galaxies and the haloes they reside within, 
which is crucial for all models of galaxy formation 
\citep[\eg][]{SAMSCole2000,Benson_SAM2003,Bower2006,DeLucia_SAM2007,Guo_SAM2013}.

The simplest and most used definition of the halo formation time, $z_f$, is given
as the epoch at which a halo's main progenitor assembles a fixed fraction
of the present day halo mass. To compute this, we follow the main progenitor branch
of each halo merger tree (see \S\ref{subsec:merger_trees} for details on the merger trees)
until the main progenitor reaches half of the halo's final mass. 
Thus, our halo formation time is the redshift of half-mass assembly, $z_{1/2}$. 
This half-mass formation time is one of the most commonly used formation time definition found 
in the literature \citep[however see][for other possible definitions]{Li_abias2008}.
Figure~\ref{fig:coco_c-M-zform} shows the formation time as a function of halo mass for our N-body
simulations. For comparison, we also show the fit to the relation obtained by \citet{MS2} for \MII{} haloes 
(rescaled from their $M_{vir}$ to our mass definition $M_{FOF}$)
and a fit to our own data. For the latter we use the same linear fit in $\log(1+z_{1/2})$ as \cite{MS2},
which takes the form:
\be
\label{eqn:m-z-fit}
1+z_{1/2} = A_0\left({M_{FOF}\over 10^{10}\Msun}\right)^{\beta}\,.
\ee
The best fit to our data was obtained for $A_0=2.77$ and $\beta=-0.0765$ and, as can be seen from the figure,
it provides a very good description of the data over the entire mass range.

Compared to the \MII{} result 
we find that our haloes formed at similar epochs, however our formation time-mass relation is steeper.
The difference can be presumably accounted for by the different cosmology and due
to different mass definitions. 
Most likely, the higher value of the $\sigma_8$ used in the \MII{} is a major driver of 
the discrepancy, as it is well established that this parameters is strongly correlated with the abundance of 
very massive haloes and hence also with the rate of their mass assembly. 

Figure~\ref{fig:coco_c-M-zform} illustrates another important point. The halo-to-halo variation in formation times
depends on halo mass, being the largest for low mass haloes and decreasing with increasing halo mass.
This trend in halo-to-halo scatter can be understood in light of {\it halo assembly bias}, which was first
pointed out by \citet{Gao_abias2007}. The halo assembly bias \citep[see also][]{Croton_abias2007,Li_abias2008}
states that the halo formation time, $z_f$, is negatively correlated with the local amplitude of the density clustering. 
It indicates that haloes form earlier in higher-density locations, which naturally have higher clustering amplitudes.
Spatial regions characterised by a higher local amplitude of the density field lead to a faster DM halo formation for 
a number of reasons: (i) because of the excess matter clustering haloes can accrete more mass in the same unit 
of time, when compared to field haloes; (ii) higher clustering amplitudes imply also higher halo merger rates;
(iii) the increased local density (compared to the universal background value) allows some density peaks to
reach more rapidly the critical density threshold, $\delta_c$, required for collapse. This reasoning can be inverted,
when applied to regions with lower clustering amplitude than the mean, where exactly the opposite
processes will induce later halo formation times. Another crucial ingredient is that DM haloes are biased tracers of the
underlying DM density field \citep[\eg][]{Frenk1985,Davis1985,Frenk1988,Cole_bias1989}, with high-mass haloes
having a large positive bias, while low mass haloes are slightly anti-biased (prefer to reside in lower-density environments).
Thus, massive haloes can only be found in regions characterised by a significant clustering excess. 
These regions are the nodes and the filaments of the cosmic web \citep[see \eg][]{Bond1996,Seth2004,Springel2006,AragonCalvo2010},
which is the most salient observational characteristic of the anisotropic nature of gravitational collapse. 
In contrast, the lower mass haloes pervade the whole range of large-scale environments, from voids to cosmic nodes,
spanning many orders of magnitude in characteristic density
\citep[for the most recent results see \eg][]{Cautun_cweb_evo_2014,Metuki2014,Falck_cweb2014,Nuza_galaxy_cweb2014}.
A detailed investigation of the origin and properties of the halo assembly/cosmic web bias down to the smallest accessible
halo mass is needed, in order to obtain a better physical understanding of this mechanism and its implication for
halo properties and galaxy formation. In principle the \coco{} simulation set, owing to its resolution and volume,
is very-well suited for such studies. However, this venture is beyond the scope of this paper and we leave it for future work.

\section{Dark Matter subhaloes}
\label{sec:subhaloes}

In hierarchical CDM cosmologies, a significant fraction of halo mass growth takes place via the accretion of lower mass haloes, 
which results in a rich substructure of orbiting smaller DM clumps called subhaloes.
The spatial distribution and abundance, kinematic and internal properties, and orbit parameters of these subhaloes are subject
of intensive study in modern cosmology. Rendering a firm insight into the various physical properties of subhaloes
plays a pivotal role in linking the observed properties of Galactic satellites and dwarf galaxy population of the Local Group
to the physical nature of dark matter. In this section we study the properties of the DM as a function of the mass of their host halo.

\subsection{Mass and velocity functions}
\label{subsec:mass_and_vel_subs}

It is well known that due to discreteness of N-body simulations, effects like over-merging, two-body
scattering, phase-space graining and force softening will affect the internal properties of haloes and subhaloes that are close 
to the resolution limit of the simulation \citep[see \eg][]{Shandarin1989,Klypin_ovm1999,Power2003,Aquarius,Abel2012,Hahn2013}. 
Among others, these effects lower the maximum velocity, $V_{max}$, of small mass objects whose $R_{max}$ is comparable to 
the gravitational force softening of the simulation. We apply the correction formula proposed by \citet[][Eqn. (10) therein]{Aquarius}, 
which, to a good approximation, under assumed perfect circular orbits, accounts for this effect. While we have done so for all our haloes and subhaloes, 
we found that, due to our very high spatial resolution, the correction has negligible effects for the vast majority of objects with $V_{max}\ge 10\rm{kms}^{-1}$ .
The subhaloes below this $V_{max}$ limit are strongly affected by numerical effects, as we discuss in detail in Appendix \ref{sec:appendix}, and are not
considered in our analysis.
\label{subsec:mas_vel_subhalo_functions}
\begin{figure}
  \includegraphics[width=85mm]{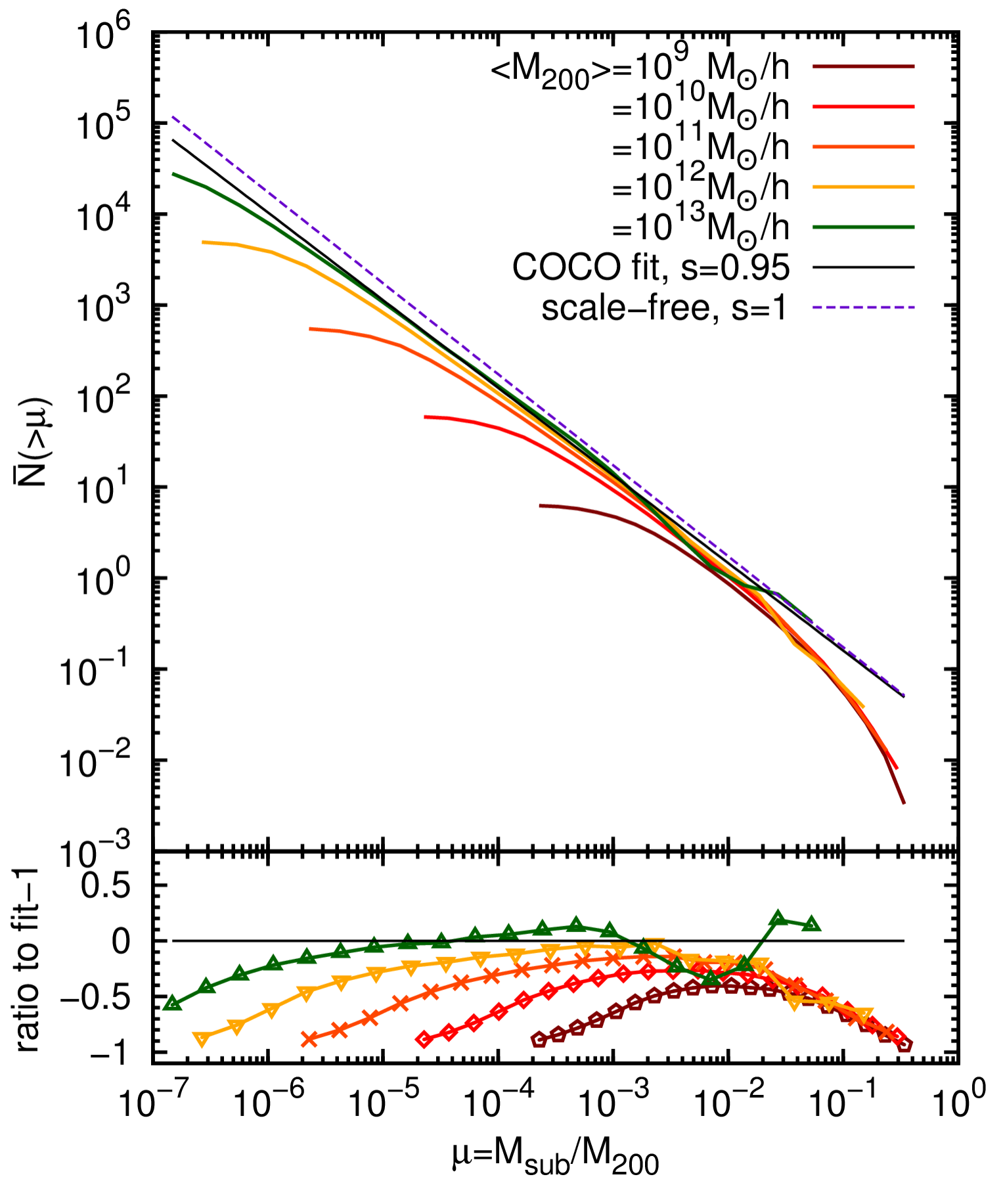}
   \caption{{\it Top panel:} The mean cumulative number of subhaloes as a function of $\mu\equiv M_{sub}/M_{200}$ binned for different host masses.
   The coloured solid lines represent hosts from a halo mass of $\langle M_{200}\rangle=10^{13}\Msun$ (green) down $\langle M_{200}\rangle=10^{9}\Msun$
   (brown). The black solid line shows the best-fitting power law function, $N(>\mu)\sim\mu^{-s}$, with $s=0.95\pm0.01$, found for the most massive host mass bin. The 
   purple dashed line illustrates the scale-free mass function with $s=1$. {\it Bottom panel:} The relative ratio of the data in each host mass bin to the \coco{} 
   best-fitting power law model with $s=0.95$.
   }
\label{fig:coco_c-subhalo_mf}
\end{figure}
\begin{table*}
\begin{minipage}{\textwidth} 
\begin{center}
\caption{The best-fitting values for the power-law index, $s$ (see Eqn.(\ref{eqn:pl_submass_func})), describing the subhalo mass function for hosts of different masses.}
\label{tab:best_fit_s}
\begin{tabular}{@{}lcccccc}
\hline\hline
$\langle M_{200}\rangle$ in $\Msun$ & $10^{10}$ & $10^{11}$ & $10^{12}$ & $10^{12}$\footnote{from {\it \aquarius{} simulation}} & $10^{13}$ & $10^{14}$\footnote{from {\it Phoenix} simulation}\\
\hline
$s$ & $0.92\pm 0.02$ & $0.93\pm0.01$ & $0.94\pm0.01$ & $0.94\pm0.02$ & $0.95\pm0.01$ & $0.97\pm0.02$\\
\hline
\end{tabular}
\end{center}
\end{minipage}
\end{table*}

Fig.~\ref{fig:coco_c-subhalo_mf} shows the mean cumulative number of subhaloes as a function of $\mu\equiv M_{sub}/M_{200}$, namely
the gravitationally bound subhalo mass in the units of its parent host halo mass. The resolution of our \coco{} run allows us to reliably identify
substructure of various size in hosts down to a halo mass of $\sim10^{9}\Msun$. To take full advantage of this,
we bin host haloes according to their mass, $M_{200}$, grouping them in five samples: $1-10\times10^9$, $1-10\times10^{10}$, 
$1-10\times10^{11}$, $1-10\times10^{12}$ and $1-1.5\times10^{13}\Msun$. 

To keep consistency with previous works and in order 
to make fair comparisons, our analysis includes all subhaloes within a radius, $r_{50}$, from the host centre, as adopted 
in the analysis of the \aquarius{} suite \citep{Aquarius}. The radius, $r_{50}$, is defined as the boundary at which the spherically averaged
density reaches a value of $50$ times the critical density for closure. On average, for galactic mass haloes, $r_{50}\simeq 1.66\times r_{200}$,
thus, one will find more subhaloes within $r_{50}$ than within $r_{200}$.
 
The \aquarius{} and Phoenix \citep{Phoenix} simulations have indicated that the substructure fractional 
mass function is well fitted over five orders of magnitude by a single power-law:
\begin{equation}
\label{eqn:pl_submass_func}
\overline{N}(>\mu)\propto\mu^{-s} \;.
\end{equation}
The best-fitting power-law suggests that $s=0.97\pm0.02$ for the Phoenix haloes and $s=0.94\pm0.02$
for the \aquarius{} suite. Both simulations have similar resolutions (for their highest level), but simulate
host halo samples of different masses. The \aquarius{} hosts have an average mass of ${\sim}10^{12}\Msun$, while the Phoenix
ones corresponds to a halo mass, $M_{200}\sim$, a few$\times 10^{14}\Msun$.

We fitted the same power-law to the substructure fractional mass function for the \coco{} hosts, to obtain the best-fitting $s$ parameter
as a function of host mass. The best fitting values and their standard error are given in Table \ref{tab:best_fit_s} and were obtained by counting
subhaloes with more than $100$ particles.
The best-fitting power-law function, for our best
resolved hosts, which have a median mass, $\langle M_{200}\rangle=10^{13}\Msun$, is shown as a solid line in Fig.\ref{fig:coco_c-subhalo_mf}.
Interestingly, this best-fitting value is found to be exactly in between the \aquarius{} and Phoenix results, with $s=0.95\pm0.01$,
but consistent with those within the fit errors.
Table \ref{tab:best_fit_s} suggests that the power-law exponent, $s$, may increase very weakly with halo mass, but, given the 
error associated with $s$, this trend is not statistically significant and a much larger study is needed to confirm or disprove such a trend.
For a closer examination, we show in the lower panel of Fig.~\ref{fig:coco_c-subhalo_mf} the fractional difference with respect to
the \coco{} best-fitting value of $s=0.95$. The panel illustrates that for all mass-binned samples there is a range in $\mu$ for which 
the fractional difference exhibits an approximately flat region. At low $\mu$, the deviations from a flat shape are driven
by numerical resolution effects, while the behaviour observed at $\mu\simgt 3\times10^{-2}$ reflects the well known
exponential cut-off in the mass function of the most massive substructures.

The best-fitting power law exponents, $s$, that we found are close to the case of a scale-free subhalo mass function with the critical value of $s=1$.
For $s=1$, each logarithmic bin in $\mu$ has an equal contribution to the total mass in subhaloes, which is logarithmically divergent as 
$\mu\rightarrow 0$. If the real substructure mass function is described by $s=1$, than a significant fraction of the host mass is contained
in subhaloes beyond the resolution limit of our simulation. For our best resolved sample with $\langle M_{200}\rangle=10^{13}\Msun$,
an average of $7.7\%$ of the host halo mass is contained in resolved substructure. Extrapolating this down to an Earth mass, corresponding to $\mu=10^{-19}$,
yields a fraction of $34\%$ mass locked in substructure. This prediction can be used further to yield DM annihilation gamma-ray flux \citep[see \eg][]{Bergstrom1998,Gondolo1999}. Detailed investigation of this situation is however beyond the scope of this paper and we leave it for the future work.

\begin{figure}
  \includegraphics[width=85mm]{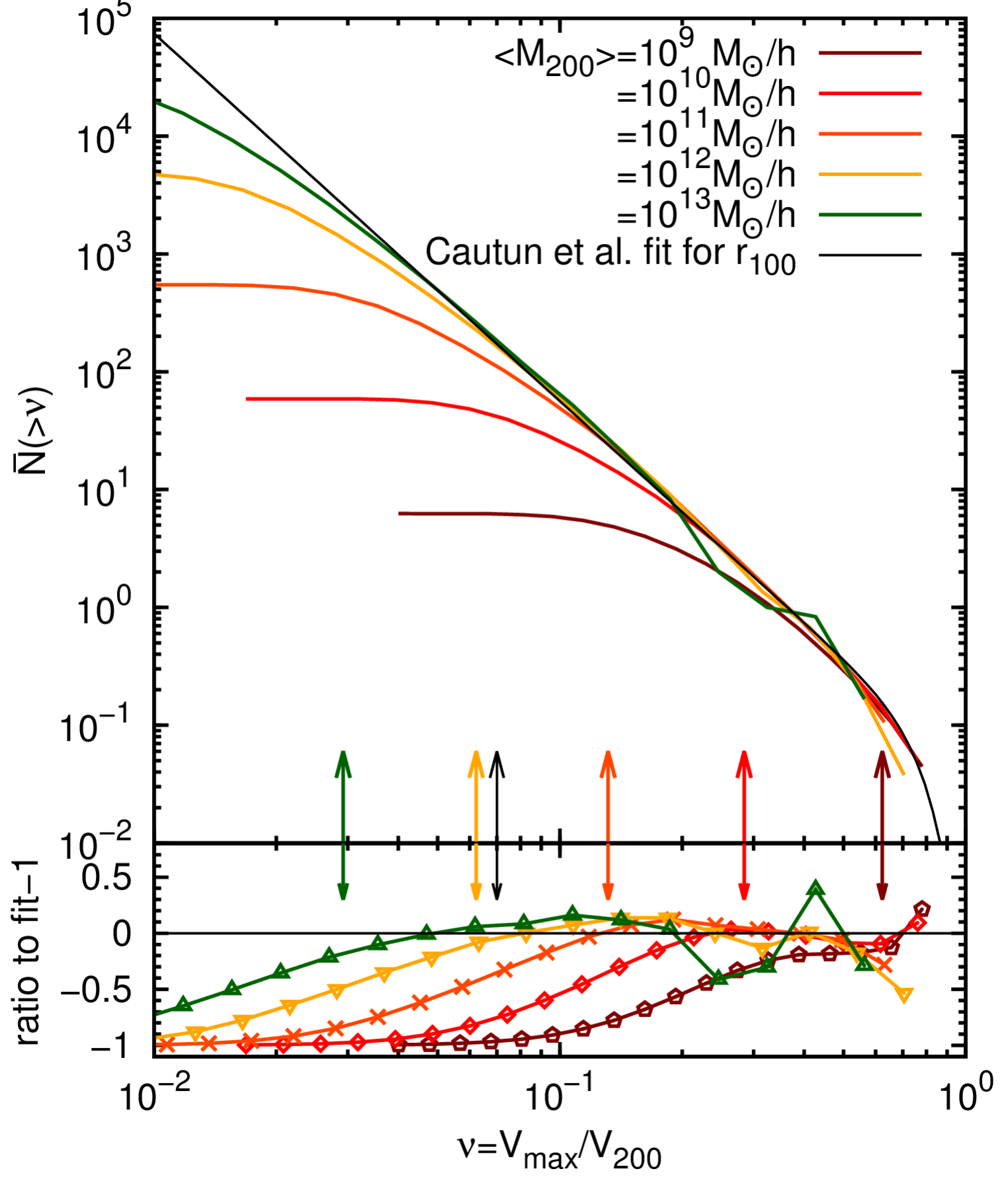}
   \caption{{\it Upper panel:} The scaled subhalo maximum circular velocity abundance as functions of the ratio $\nu=V_{max}/V_{200}$ 
   for hosts of different masses.
   The vertical arrows mark the $\nu$ value corresponding to $V_{max}=10$ kms$^{-1}$, which is our resolution limit.
   The black thin line shows the best-fitting function of \citet{Cautun2014b} for subhaloes within a radius, $r_{100}$, 
   from the host. {\it Lower panel:} Ratios between the mean subhalo count at various host masses and the Cautun et al. result.
   }
\label{fig:coco_c-subhalo_Nu-mf}
\end{figure}
\begin{figure}
  \includegraphics[width=85mm]{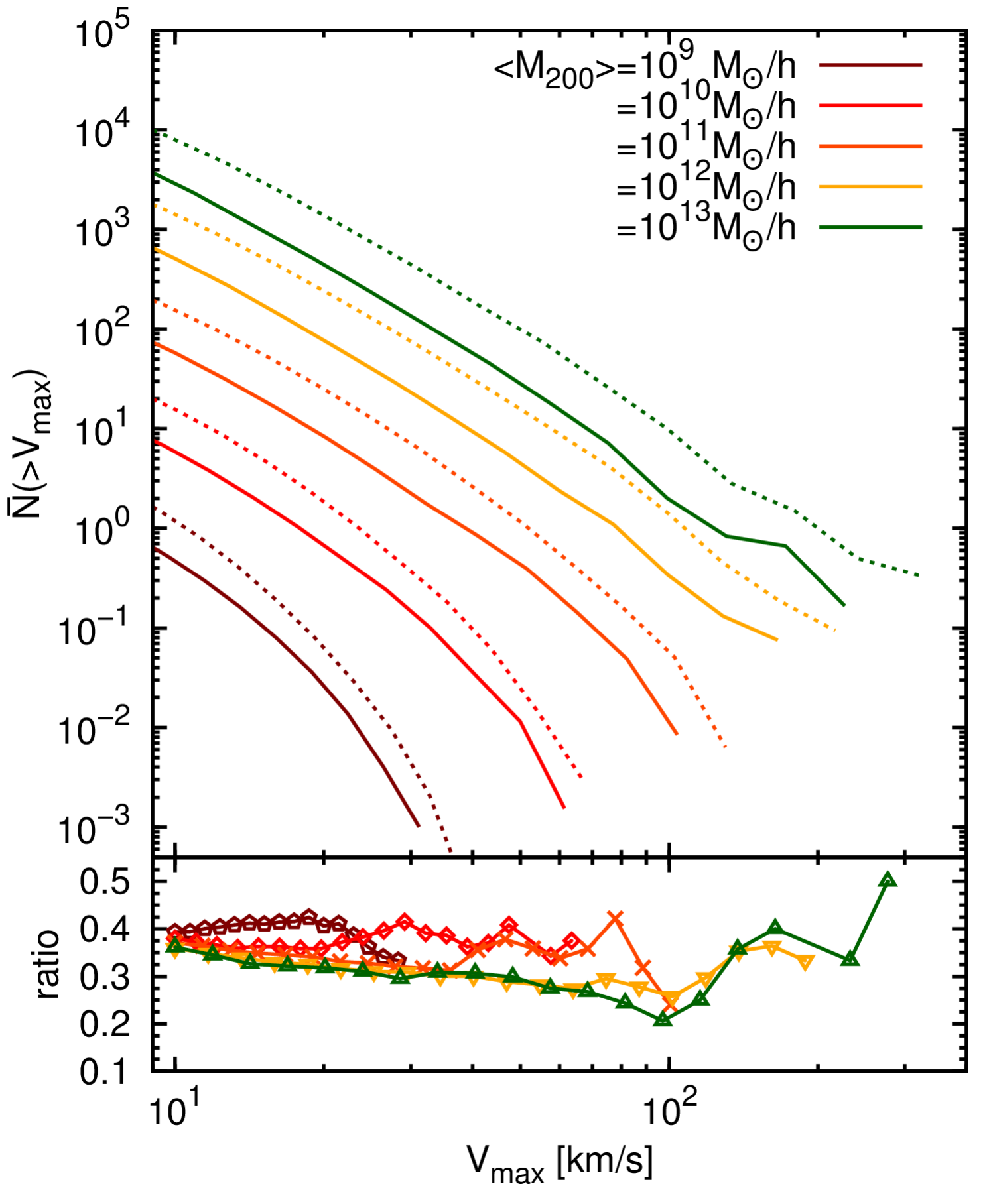}
   \caption{{\it Upper panel:} The subhalo count as a function of maximum circular velocity for host haloes of different masses.
   The solid lines show the subhalo count using the present ($z=0$) $V_{max}$ values. The dotted lines depict the subhalo count
   using the $V_{max}$ value at the subhalo's infall time. {\it Lower panel:} The ratio between the subhalo count
   using present time $V_{max}$ and the subhalo count using the $V_{max}$ at infall time. }
\label{fig:coco_c-subhalo_vmaxfun}
\end{figure}

It has been long postulated that the scaled subhalo velocity function is independent of host halo mass,
when expressed as a function of $\nu=V_{max}/V_{200}$, 
which is the ratio between the subhalo maximum circular velocity and the host virial velocity \citep[e.g.][]{Moore1999,Kravtsov2004}.
Recently, this has been thoroughly confirmed using a large number of host haloes \citep{Wang2012,Cautun2014b}.
Given both the very high resolution and the large number of haloes in the \coco{} simulation, we can investigate the
postulated invariance of the scaled subhalo velocity function over a wider dynamical range in subhalo $V_{max}$ and down
to lower host halo masses. This is shown in Fig.~\ref{fig:coco_c-subhalo_Nu-mf}, where we plot the mean cumulative 
satellite count, $\overline{N}(>\nu)$, as a function of $\nu$ for hosts binned according to their halo mass.
To better highlight the invariance with host halo mass, the lower panel of Fig.~\ref{fig:coco_c-subhalo_Nu-mf}
shows the ratio between $\overline{N}(>\nu)$ measured in \coco{} and the best fit of \cite{Cautun2014b} for the mean 
subhalo count around galactic mass haloes. Since \cite{Cautun2014b} does not compute the subhalo count within $r_{50}$,
we take their result for subhaloes found within a distance of $r_{100}$ from the centre of the host halo. 
This leads to us counting more subhaloes than \cite{Cautun2014b}, which explains why the \coco{} results are systematically 
above the zero level in the bottom panel of the figure. 

We find that the mean subhalo count, $\overline{N}(>\nu)$, exhibits at most a very weak 
dependence on host mass. This should be compared to the satellite abundances of Fig.~\ref{fig:coco_c-subhalo_mf},
which show a strong dependence on host mass, $M_{200}$. Any systematic deviations from a flat line for the results shown
in the bottom panel of Fig.~\ref{fig:coco_c-subhalo_Nu-mf} appear only below the resolution limit of the simulation,
which is show by a vertical arrow for each halo mass bin. 
The deviations seen at high $\nu$ values for the most massive bin, $M_{200} = 10^{13}\Msun$,
are due to the small number of host haloes present in that sample and hence are not significant.
Thus, our results confirm the postulated invariance
of $\overline{N}(>\nu)$ over four orders of magnitude in host mass, showing that this assumption
holds for the majority of DM haloes that can host galaxies ($10^{10}\simlt M_{200} h/M_{\odot}\simlt 10^{13}$). 
This invariance was exploited by \citet{Wang2012} and \citet{Cautun2014a} to derive new theoretical constrains 
on the mass of the Milky Way halo.

In Fig.~\ref{fig:coco_c-subhalo_vmaxfun} we show the mean subhalo count, $\overline{N}(>V_{max})$, as a function of
subhalo maximum velocity, $V_{max}$. Describing subhaloes in terms of the maximum circular velocity is more closely 
related to observations, since $V_{max}$ is more easily measured in observations. 
The figure compares the subhalo abundance using the present day $V_{max}$ as well as the subhalo
count as a function of the maximum circular velocity at subhalo's infall time, $V_{max}^{inf}$.
The $V_{max}^{inf}$ values are obtained by tracing the merger tree of each subhalo and taking the peak value of $V_{max}$
throughout the history of the subhalo. For most practical applications, $V_{max}^{inf}$ is well approximated
by the peak value of $V_{max}$, since, once a halo falls into a more massive object,
it becomes the subject of intensive tidal stripping and so its $V_{max}$ value is very likely to decrease rather
than increase. Fig.~\ref{fig:coco_c-subhalo_vmaxfun} shows
that at fixed subhalo size, i.e. fixed $V_{max}$ values, the abundance of objects roughly
increase by an order of magnitude for each order of magnitude in host mass. This scaling is most pronounced for
sufficiently small objects. This scaling breaks down for the most massive subhaloes, since the subhalo abundance 
changes its shape from a power-law to an exponential decline (see Fig.~\ref{fig:coco_c-subhalo_mf} 
and ~\ref{fig:coco_c-subhalo_Nu-mf}). 
Interestingly, a similar scaling is found also for the subhalo abundance as a function of $V_{max}^{inf}$. This can
readily be seen from the bottom panel Fig.~\ref{fig:coco_c-subhalo_vmaxfun} that shows the ratio
$\mathcal{R}=\overline{N}(>V_{max})/\overline{N}(>V_{max}^{inf})$. Both these functions are calculated using
the same objects, found at $z=0$ inside a distance, $r_{50}$, from their host, so their values are not influenced by the
destruction or accretion of new subhaloes. In other words, we expect that the ratio $\mathcal{R}$ 
and its departure from unity are a good proxy for the efficiency of subhalo tidal striping at fixed subhalo circular velocity.
For small subhaloes with $V_{max}\simlt 20$kms$^{-1}$, we find that the ratio approaches $\mathcal{R}\approx 0.35$ for all host masses except
for the lowest mass bin. For this lowest mass sample, $\langle M_{200}\rangle=10^{9}\Msun$, \coco{} has enough resolution 
to identify only the most massive subhaloes and it does not capture the power-law like regime of the subhalo abundance function.
The convergence of the $\mathcal{R}$ ratios towards a single values indicates that the efficiency of subhalo tidal stripping 
is comparable in hosts that differ by four orders of magnitude in mass, provided that the considered subhaloes are sufficiently small
in comparison to their host.

\subsection{The radial distribution}
\label{subsec:sub_radial_dist}
\begin{figure*}
  \includegraphics[width=170mm]{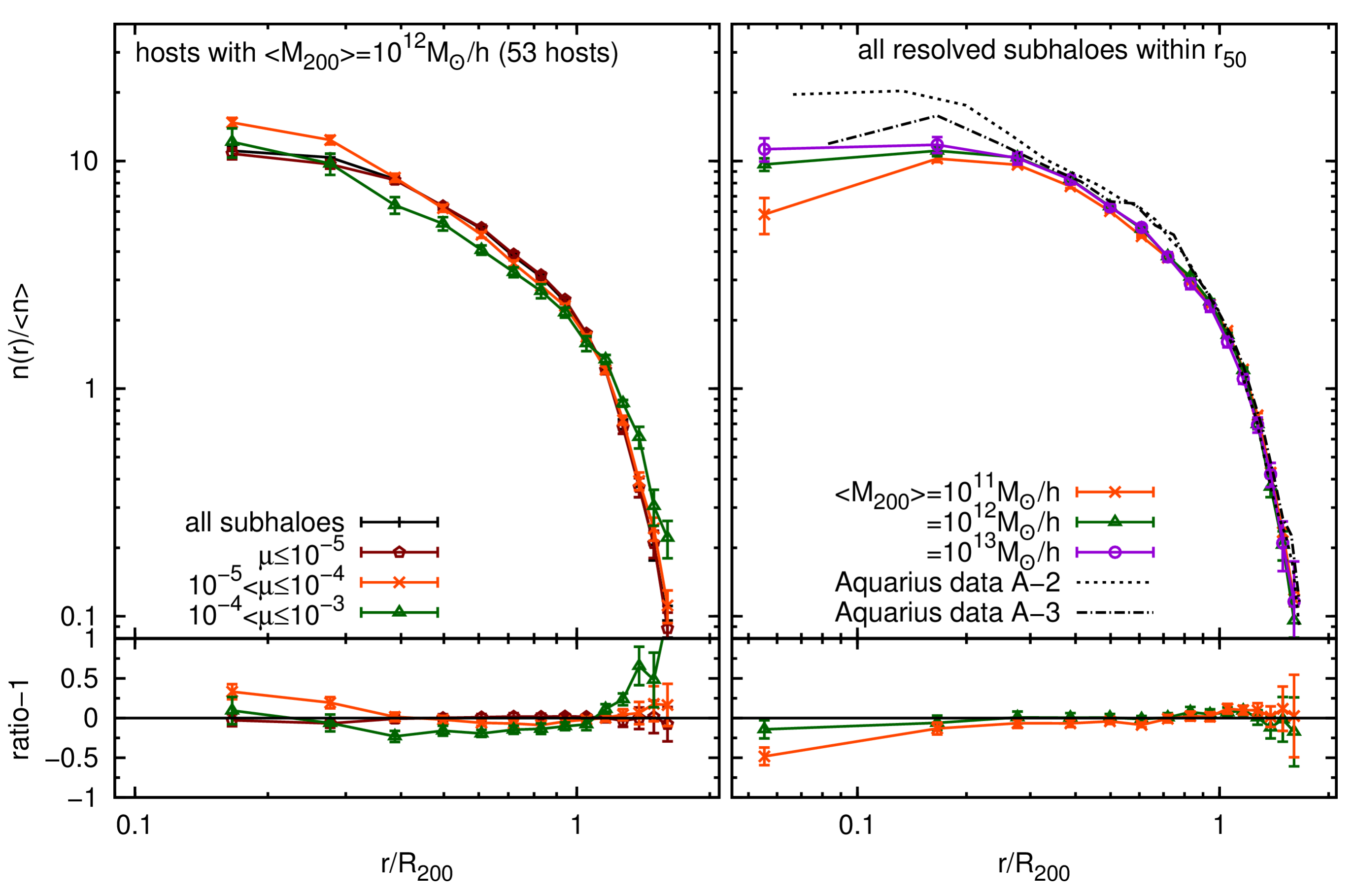}
   \caption{Radial number density profiles of subhaloes. {\it Left panel:} The radial distribution of subhaloes in galactic mass hosts, i.e 
   $\langle M_{200}\rangle=10^{12}\Msun$. The lines with symbols corresponds to subhaloes of various masses,  
   as expressed in terms $\mu=M_{sub}/M_{200}$, while the black solid
   line with crosses shows the results for all subhaloes. The bottom panel give the ratio of the various subhalo mass bins
   with respect to the distribution of all subhaloes.
   {\it Right panel:} The radial distribution of subhaloes for different host masses.
   The lines with symbols show the results of \coco{}. The dotted line (dashed-dotted) show the equivalent 
   substructure profile from the \aquarius{} A-2 (A-3) run. The bottom panel shows the ratio of each data set to the reference case, 
   which we take as $\langle M_{200}\rangle=10^{13}\Msun$.
   }
\label{fig:coco_c-subhalo_raddist}
\end{figure*}

The radial distributions of subhaloes is also a subject of intensive study \citep[\eg][]{Gao2004,Diemand2004,DeLucia2004,Nagai2005,Wang2013}, 
since understanding how DM substructures are distributed inside their host haloes is important for several reasons. 
Among others, the radial distribution of subhaloes is instrumental in connecting the observations of satellite 
galaxies with the properties of the background cosmology; it serves as input for many semi-analytical 
galaxy formation models; and it is important for strong lensing studies 
\citep[\eg][]{Mao1998,Metcalf2001,Metcalf2002,Kochanek2004,Dandan2015}.

\citet{Aquarius} have found that the radial distribution of subhaloes is independent of subhalo mass for at least
five decades in mass (see Fig.~11 therein). We further investigate this finding, since the large number of host haloes 
of the \coco{} run allow for much better statistics. In addition, we further extended the analysis of Springel \etal{} by
studying how the radial distribution of subhaloes varies with host mass.

The left-hand panel of Fig.~\ref{fig:coco_c-subhalo_raddist} shows the dependence of the subhalo radial distribution on subhalo mass. The subhalo
population is split according to their rescaled mass, $\mu=M_{sub}/M_{200}$, following which, we
stack the radial profiles of 53 host haloes whose median mass is $\langle M_{200} \rangle = 10^{12}\Msun$. 
In the bottom-left panel we show the ratio of each subhalo mass sample with respect to the reference 'all subhaloes' line. 
We find a large degree of self-similarity between subhaloes of different masses, in agreement with the results of \citet{Aquarius}.
However, we do find a weak, but systematic trend with subhalo mass. This is especially pronounced
for the most massive subhaloes with $\mu>10^{-4}$ for which, when compared to the distribution of all subhaloes, the radial distribution 
has an excess for $r>r_{200}$ and a scarcity at smaller radii. The exact size of this systematic effect is difficult to pinpoint because of the relatively
large uncertainties associated with our data, which are caused by a significant host-to-host scatter in the distribution of massive subhaloes.

The top-right panel gives the radial distribution of all subhaloes for various host halo masses. 
In addition, we also show the results of the \aquarius{} A-2 and A-3 runs, to find that
the \coco{} data for the best resolved haloes of mass $\langle M_{200} \rangle=10^{13}\Msun$ agrees with the \aquarius{} results
down to a radial distance of $\sim 0.3 r_{200}$. Below that radius, \coco{} contains fewer resolved subhaloes than the higher resolution
\aquarius{} runs. A similar behaviour is seen when comparing the A-3 results to the A-2 ones, which is its higher resolution counterpart, and also
when comparing the $\langle M_{200}\rangle=10^{11}\Msun$ sample to the $\langle M_{200}\rangle=10^{13}\Msun$ one. 
The systematic difference between \coco{} and \aquarius{} at $r\sim0.6R_{200}$ is likely a manifestation
of the fact that the \aquarius{} sample contains a single halo and hence it is an indication of the object-to-object scatter.
From the bottom-left panel of Fig.~\ref{fig:coco_c-subhalo_raddist}, which shows the ratio with respect to the reference 
$\langle M_{200} \rangle=10^{13}\Msun$ sample, we find that the radial distribution agrees remarkably well down to $r\sim 0.2 r_{200}$ 
for hosts spanning three decades in mass ($10^{11}-10^{13}\Msun$). 
We can conclude that the spatial distribution of low mass satellites (\ie with $\mu\ll 10^{-3}$) has a universal shape across hosts of different masses. 
This is yet another example of the self-similar character of DM haloes in CDM cosmologies. Detailed analysis of the subhalo radial density profiles 
is beyond the scope of our current paper and we leave it for a future work.

\subsection{The \ensuremath{V_{max}-R_{max}} relation}
\label{subsec:vmax-rmax}
\begin{figure}
  \includegraphics[width=85mm]{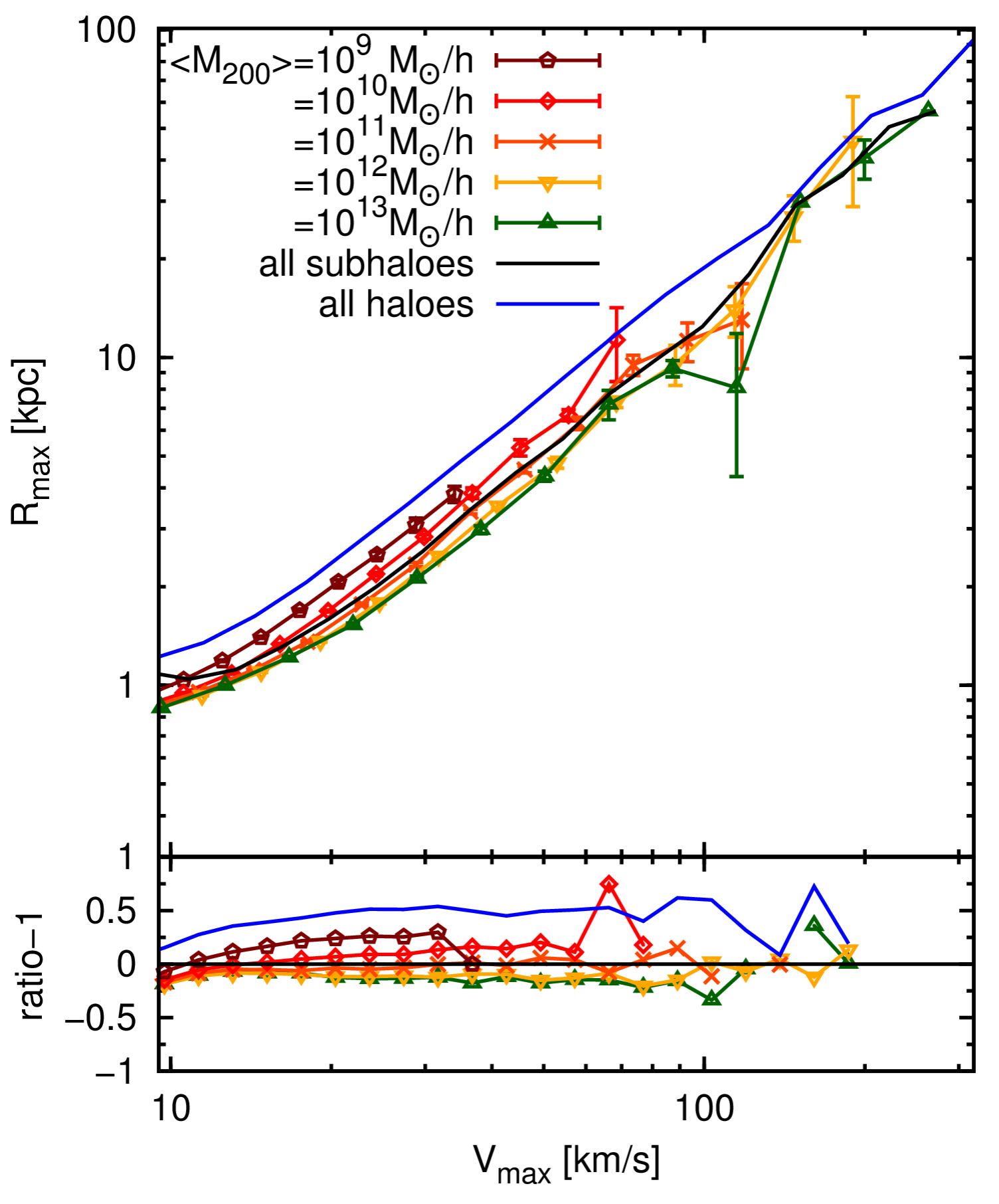}
   \caption{The $V_{max}-R_{max}$ diagram for subhaloes and haloes. 
{\it Upper panel:} the relation between the
maximum circular velocity and the radius at which it is attained for subhaloes found in hosts of various masses
(solid lines with symbols and error bars). The solid black line shows this relation for all subhaloes in the simulation,
while the solid black line illustrates this relation for field haloes. {\it Lower panel:} the ratios of 
the different samples with respect to the 'all subhaloes' sample.
}
\label{fig:coco_c-subhalo_vmax_rmax_diffhosts}
\end{figure}

A fundamental structural property of subhaloes is the relation between the maximum circular velocity, 
$V_{max}$, and the radius corresponding to this maximum, $R_{max}$. 
For DM dominated objects like spheroidal dwarf galaxies,
the kinematics of the stellar component can be related to the underlying DM density profile
via the ${V_{max}-R_{max}}$ relation, which is the basis of numerous cosmological studies based on
the stellar kinematics of Local Group dwarf galaxies
\citep[see \eg][]{2011MNRAS.415L..40B,2012MNRAS.422.1203B,Sawala2013,Sawala2014a,Sawala2014b,Sawala2014c,Wang2012,Cautun2014a,DiCinto2013,ZOlotov2012,Arraki2014}.
While it is not our intention to have a detailed study of the $V_{max}-R_{max}$ relation, we would like to add to the discussion
by presenting an interesting finding. Namely, in Fig.~\ref{fig:coco_c-subhalo_vmax_rmax_diffhosts}
we plot the $V_{max}-R_{max}$ diagram for both haloes and subhaloes. The lines with error bars mark the mean
relation for subhaloes found in hosts of different masses. The solid black line corresponds to the values obtained
by considering all resolved subhaloes in our simulation, while the solid blue curve indicates the same relation computed
for field, isolated haloes. The mean $R_{max}$ value at fixed $V_{max}$ is a crude measure of central (sub)halo
densities, as objects with the same $V_{max}$, but larger (smaller) $R_{max}$ values are characterised by lower (higher) central
densities. Fig.~\ref{fig:coco_c-subhalo_vmax_rmax_diffhosts} shows that at fixed $V_{max}$, the mean $R_{max}$ values
for haloes are $50\%$ higher than for subhaloes. This can be easily seen as the blue solid line in the bottom panel of the figure.
This behaviour reflects a well known fact, that subhaloes tend to have more concentrated density profiles due 
to the effects of tidal stripping, which become significant once the subhaloes fell into their respective hosts 
\citep[see \eg][]{Moore1999,ViaLactea,Aquarius}
The tidal forces truncate a subhalo's density profile by removing the mass that is only weakly gravitationally bound to the object.
Since field haloes are rarely the subject of severe tidal forces, no such stripping takes place.
An even more interesting find is the systematic difference, at fixed $V_{max}$, between the mean $R_{max}$ values 
characterizing subhaloes found in host haloes of different masses.
Hence, satellites with the same $V_{max}$ values tend to have systematically higher $R_{max}$ values
when found in central haloes of lower mass.

\section{Conclusions}
\label{sec:conclusions}

Since the establishment in the late 80s and 90s of the CDM model as the standard model for cosmic structure formation, it has
become a subject of extensive tests and scrutiny. To understand the process of galaxy formation and evolution in a hierarchical 
$\lcdm$ cosmology, we need detailed knowledge of a multitude of physical processes that act over an overwhelming range of scales. 
The formation and dynamical evolution of haloes and subhaloes, from tiny DM specks of Earth mass to the most massive gravitationally bound objects,
together with highly non-linear and complicated baryonic processes set the framework in which galaxies live and evolve in our Universe.
The constant development of observational techniques is calling for an improvement in our theoretical modelling and understanding of the crucial 
phenomena involved.
For this reason, we need simulations with ever growing resolution. However, we also need to model large enough cosmic volumes to
obtain reliable statistics for various objects, from dwarf to giant galaxies. This is where simulations like the {\it Copernicus Complexio} play
a pivotal role, since they have both a very high resolution and a large cosmological volume. In this paper we have introduced
a new simulation, the \coco{}, that can reliable resolve all substructure down to a $V_{max}\sim 10$ kms$^{-1}$ in a cosmological relevant
volume of $\sim 2.2\times 10^4\hmpcc$. This simulation is the first of its kind and is meant to be part of a whole series
of intermediate zoom-in runs implementing both more $\lcdm$ cosmic volumes but also alternative DM physics like Warm or Self-Interacting DM models
\citep[see also][]{Bose2016}. 

The following is a summary of our main results:
\begin{itemize}
 \item The FOF mass function matches the ST and Reed formulas over seven orders of magnitude in halo mass at $z=0$. However, for the intermediate redshift range
       of $2<z<0.5$, the ST formula tends to over-predict the number of collapsed objects.
 \item We have observed a departure of the $c-M_{200}$ relation from a single power-law at lower halo masses, in agreement with the 
       results of \citet{Sanchez-Conde2014}.  We give a best fit to the \coco{} data that reliably describes the concentration-mass relation of relaxed 
       haloes over six decades in halo mass $10^8\leq M_{200}/(\Msun)\leq10^{14}$.
 \item We have probed the redshift evolution of the $c-M_{200}$ relation in the redshift interval, $9\leq z \leq 0$, to find that it is monotonic 
       for small halo masses.
 \item The hierarchical nature of halo formation processes is confirmed for seven orders of magnitude in mass. The object-to-object scatter
       of the halo formation time depends on halo mass, with lower mass haloes showing a significantly larger scatter. This most likely
       is a manifestation of halo assembly bias, reflecting the multitude of environments in which low mass haloes are formed and evolve.
 \item We have confirmed the power-law character of the subhalo mass function, $\overline{N}(>\mu)\propto \mu^{-s}$, down to a rescaled subhalo mass, $\mu=10^{-6}$.
       For our best resolved hosts, with median halo mass, $\langle M_{200} \rangle=10^{13}\Msun$, we find a power-law exponent, $s=0.95\pm0.01$.
 \item We find that the power-law exponent, $s$, depends on the host halo mass. It varies from $s=0.97\pm0.02$ for cluster mass haloes \citep{Phoenix} 
       to $s=0.92\pm0.02$ for $10^{10}\Msun$ haloes. 
 \item Our data confirms over a wider dynamical range in subhalo sizes and down to lower host masses that the mean subhalo abundance,
        $\overline{N}(>\nu)$, when expressed in terms of $\nu=V_{max}/V_{200}$, is to a very good approximation independent of host halo mass.
       The best-fitting results for $\overline{N}(>\nu)$, which were proposed by \cite{Cautun2014b}, match our data down to our resolution limit.
 \item The radial distribution of galactic subhaloes is nearly independent of subhalo mass, albeit with a very weak trend. Due to a large host-to-host
      scatter, this trend becomes visible only once we average over a substantial number of host haloes. In addition, the radial distribution of subhaloes
      is nearly universal for hosts differing by three orders of magnitude in halo mass.
 \item Finally, we have found that at fixed $V_{max}$ the mean $R_{max}$ values of subhaloes depend on the host halo mass, 
       with lower mass hosts having subhaloes with higher $R_{max}$ values.       
       This most likely reflects that at fixed subhalo size the tidal stripping processes are more efficient in more massive hosts.
\end{itemize}

The current and future runs of the \coco{} suite will allow us to further test models of cosmic structure formation, including 
the development of semi-analytical
galaxy formation models into the regime of low mass (sub)halo 
(hence also low galaxy luminosity) \citep[for more details see][]{GuoCOCOSAM}. This new satellite galaxy catalogue
build on the base on \coco{} was alreay used for stringent statistical study of the prevalence of rare plannar sattellite configurations
in the $\lcdm$ \citep{CautunRotSat}.
Moreover, our new set of simulations will allow for a better statistical study of radio-flux anomalies and lensing arc-distortions, 
the low-luminosity galaxy population, reionisation treatment in semi-analytical models and effects of the large-scale structures 
(Cosmic Web) on (sub)halo and galaxy properties and distributions. As these projects are currently work in progress, with 
the publication of this paper we also intend to make publicly available, in a short time, all relevant \coco{} halo and subhalo data bases accompanied
by semi-analytical galaxy catalogues. In doing so, our hope and intention is to allow other researchers to use the \coco{} data for their own research projects.

\section*{Acknowledgements}
We thank the anonymous referee for valuable comments that helped improve the scientific quality of this manuscript.
The authors are very grateful to Alex Knebe and Volker Springel for their comments and support at the early stages of this project.
We are very grateful to Aaron Ludlow, Jaxin Han, Shaun Cole, Matthieu Schaller, Qi Guo and Julio Navarro 
for various suggestions and interesting discussions that help to increase the scientific
value of this paper.
We would like to acknowledge Lydia Heck of Durham University and Aleksander Niegowski of University of Warsaw 
for their technical support and invaluable help during the run and analysis of our simulations.
WAH, CSF and MC thank the ERC Advanced Investigator grant COSMIWAY [grant number GA 267291] and 
the Science and Technology Facilities Council [grant number ST/F001166/1, ST/I00162X/1].
WAH was also partially supported by the Polish National Science Center under contract \#UMO-2012/07/D/ST9/02785.
This work used the DiRAC Data Centric system at Durham University, operated
by ICC on behalf of the STFC DiRAC HPC Facility (www.dirac.ac.uk). This equipment was funded by BIS National
E-infrastructure capital grant ST/K00042X/1, STFC capital grant
ST/H008519/1, and STFC DiRAC Operations grant ST/K003267/1 and Durham
University. DiRAC is part of the National E-Infrastructure.  This
research was also carried out with the support of the ``HPC Infrastructure
for Grand Challenges of Science and Engineering'' Project, co-financed
by the European Regional Development Fund under the Innovative Economy
Operational Programme.

\appendix
\section{Numerical convergence and resolution test}
\label{sec:appendix}
Here we assess a conservative limit for the mass and maximum circular velocity of the haloes and subhaloes that were 
resolved reliably in our simulations. This is necessary since haloes and especially subhaloes that are resolved at low resolution
are subject to many numerical artefacts that can alter their inner properties like density and circular velocity profiles,
leading to unphysical results. In the following, we present two tests for determining the resolution limit of our numerical experiment.

\begin{figure}
  \includegraphics[width=85mm]{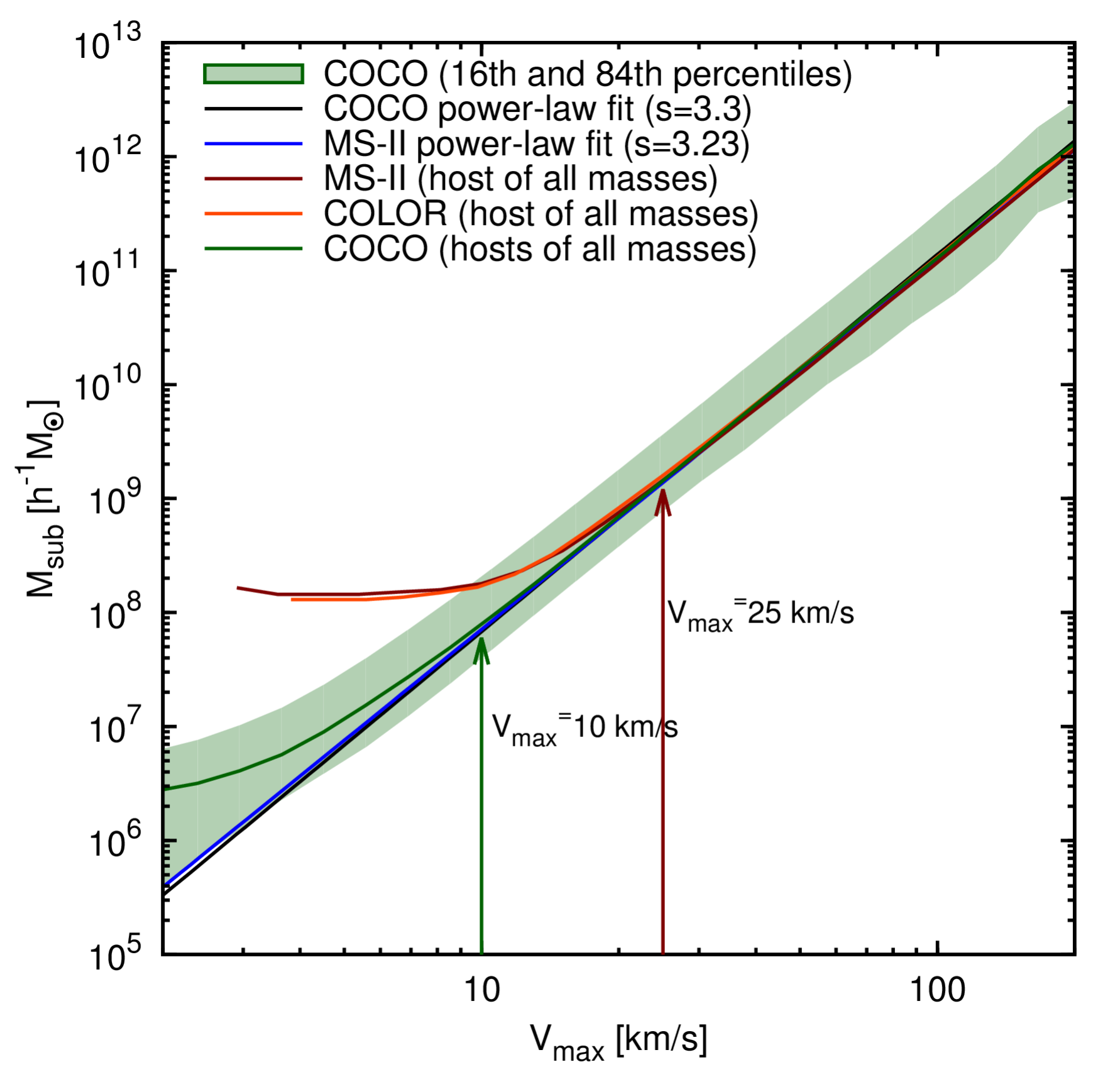}
   \caption{A comparison of the $V_{max}-M_{sub}$ relation for subhaloes found in hosts of all masses at redshift, $z=0$.
    The \coco{}, \scolor{} and \MII{} lines depict the median relation found by binning the subhaloes according to their $V_{max}$ values.
    The shaded green region illustrates the 16th to 84th percentiles around the median for the \coco{} sample. We also show the best 
    fit power laws (Eq. \ref{eqn:vmax-msub}) to the \coco{} and \MII{} data.}
\label{fig:coco_conv1}
\end{figure}

The first test investigates the relationship between the mass, $M_{sub}$, and the maximum circular velocity, $V_{max}$, of subhaloes. 
Following \citet{Boylan-Kolchin2010}, we fit the relation with the power-law
\be
\label{eqn:vmax-msub}
M_{sub}= 6\times 10^{10} \left({V_{max}}\over {100 \textrm{km s}^{-1}}\right)^{s}\,\Msun\,.
\ee
\citeauthor{Boylan-Kolchin2010} have found that such a power law, with $s=3.23$, provides a very good description of the 
\MII{} data down to $V_{max}=25$ km s$^{-1}$. Below this value, the $V_{max}-M_{sub}$ relation deviates from the power law
fit suggesting that subhaloes with lower $V_{max}$ values are affected by numerical resolution effects. The same power law,
albeit with a slightly steeper scaling exponent, $s=3.3$, gives a very good fit to the \coco{} data too, as shown in
Fig.~\ref{fig:coco_conv1}. The \coco{} subhaloes follow the power law relation down to a much smaller $V_{max}$ 
values of $\sim10$ km s$^{-1}$ ($M_{sub}\sim5\times10^7\Msun$). 
Fig.~\ref{fig:coco_conv1} also shows that that there is a very good convergence between the
\scolor{} and the \coco{} runs, with \scolor{} having a resolution limit of $V_{max}=25$ km s$^{-1}$. \scolor{} has the
same behaviour as \MII{} since both simulations have the same particle mass and force resolution. 
\begin{figure}
  \includegraphics[width=85mm]{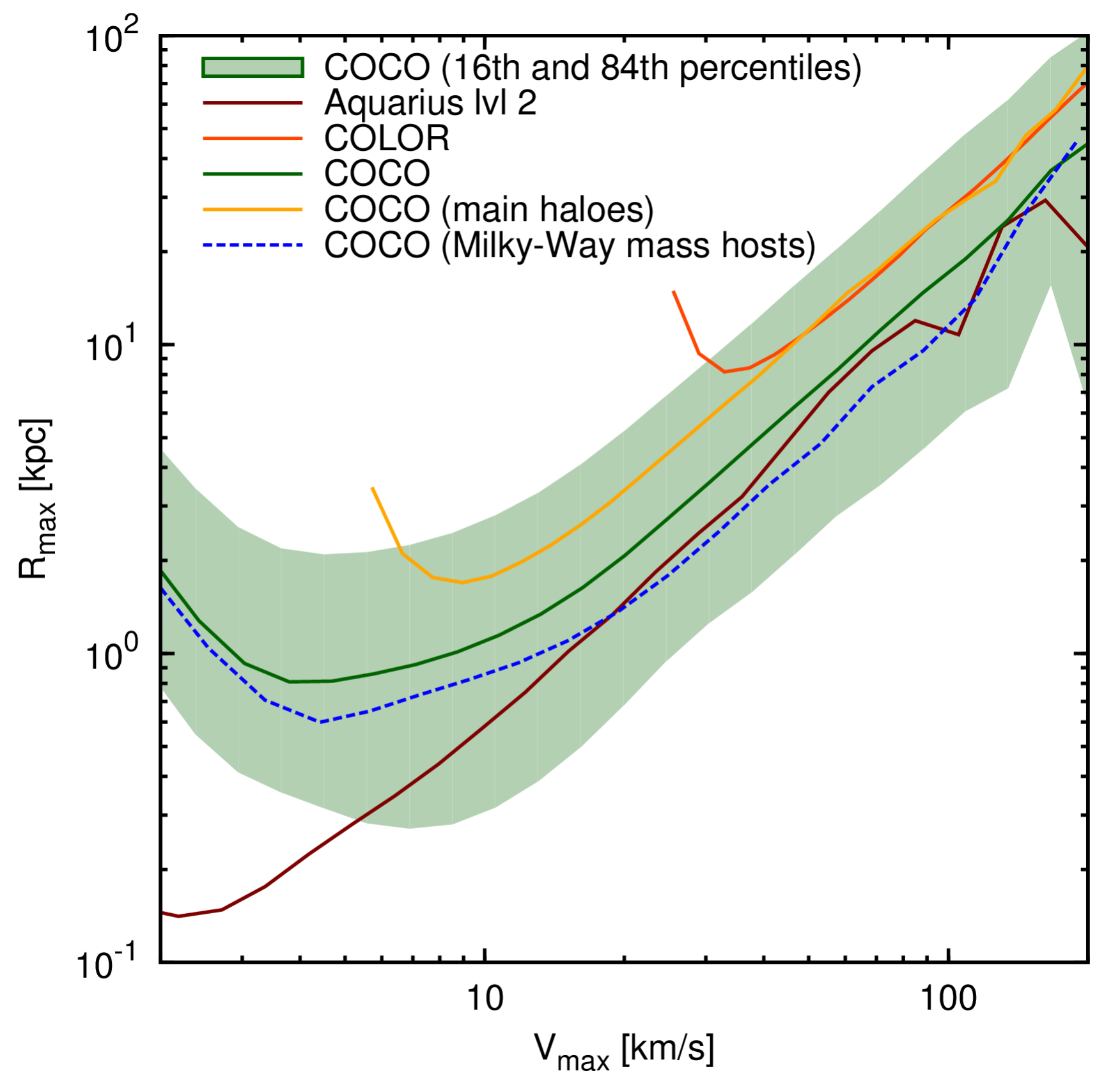}
   \caption{A comparison of the $V_{max}-R_{max}$ relation for haloes versus that for subhaloes. The solid yellow and orange
    lines show the relation for main haloes in the \coco{} and \scolor{} samples. The remaining lines show the $V_{max}-R_{max}$ 
    relation for subhaloes identified in all the \coco{} hosts (green solid line); in Milky-Way mass hosts 
    ( blue dashed line;see \S\ref{subsec:vmax-rmax}); and in the \aquarius{} level 2 hosts (solid brown line). 
    The green shaded region gives the 16th to 84th percentiles for the full sample of \coco{} subhaloes. }
\label{fig:coco_conv2}
\end{figure}

In the second test we compare the $V_{max}-R_{max}$ relation of subhaloes with the same relation for haloes (see Fig.~\ref{fig:coco_conv2}).
We find that both for haloes and subhaloes, the median $V_{max}-R_{max}$ relation shows an upturn indicative of numerical
resolution effects at $V_{max}=10$ and 25 km s$^{-1}$ for \coco{} and \scolor{}, respectively. Thus, the two $V_{max}$
thresholds give a good conservative estimate for the resolution limit of our two simulations.

\bibliographystyle{mn2e}
\bibliography{coco_intro}

\bsp

\label{lastpage}

\end{document}